\documentclass[preprint,prd,aps,showpacs,showkeys,nofootinbib]{revtex4}
\usepackage{graphicx}
\begin{document}

%\preprint{hep-ph/0412147}

\title{Gauged baryon and lepton numbers in supersymmetry with a $125\;{\rm GeV}$ Higgs}

\author{Tai-Fu Feng$^{a}$\footnote{email:fengtf@hbu.edu.cn}, Shu-Min Zhao$^a$\footnote{email:smzhao@hbu.edu.cn},
Hai-Bin Zhang$^{a,b}$, Yin-Jie Zhang$^a$, Yu-Li Yan$^a$}

\affiliation{$^a$Department of Physics, Hebei University, Baoding, 071002, China\\
$^b$Department of Physics, Dalian University of Technology,
Dalian, 116024, China}

\begin{abstract}
Assuming that the Yukawa couplings between the Higgs and exotic quarks cannot be ignored,
we analyze the signals of decay channels $h\rightarrow\gamma\gamma$ and $h\rightarrow VV^*\;(V=Z,\;W)$
with the Higgs mass around $125\;{\rm GeV}$ in a supersymmetric extension of the standard model where baryon
and lepton numbers are local gauge symmetries. Adopting some assumptions on relevant parameter space,
we can account for the experimental data on Higgs from ATLAS and CMS naturally.
\end{abstract}


\maketitle

\section{Introduction\label{sec1}}
\indent\indent
The main destination of the Large Hadron Collider (LHC) is to understand the origin of the electroweak symmetry
breaking, and searches the neutral Higgs predicted by the standard model (SM) and its various extensions.
Recently, ATLAS and CMS have reported significant excess events which are interpreted probably to be
related to the neutral Higgs with mass $m_{h_0}\sim 124-126\;{\rm GeV}$\cite{CMS,ATLAS}. 
This implies that the Higgs mechanism to break electroweak symmetry possibly has a solid 
experimental cornerstone.

As the simplest soft broken supersymmetry theory, the minimal supersymmetric extension of the standard model
(MSSM) \cite{MSSM} has drawn the attention from physicist for a long time. Furthermore,
Broken baryon number (B) can explain the origin of the matter-antimatter asymmetry in the Universe naturally.
Since heavy majorana neutrinos contained in the seesaw mechanism can induce the tiny neutrino masses\cite{seesaw}
to explain the neutrino oscillation experiment, lepton number (L) is also expected to be broken.
Ignoring Yukawa couplings between Higgs doublets and exotic quarks, the authors of literature\cite{BL_h, BL_h1}
investigate the predictions for the mass and decays of the lightest CP-even Higgs in a minimal local
gauged B and L supersymmetric extension of
the SM which is named BLMSSM. Since the new quarks are vector-like, one obtains that 
their masses can be above $500\;{\rm GeV}$ 
without assuming large couplings to the Higgs doublets in this model. 
Therefore, there are no Landau poles for the Yukawa couplings here.
Additionally, literature\cite{BL} also examines two extensions of the SM
where $B$ and $L$ are spontaneously broken gauge symmetries around ${\rm TeV}$ scale.
Assuming that the Yukawa couplings between Higgs and exotic quarks cannot be ignored here,
we investigate the lightest CP-even Higgs decay channels $h\rightarrow\gamma\gamma$, $h\rightarrow VV^*\;(V=Z,\;W)$
in the BLMSSM.

Our presentation is organized as follows. In section \ref{sec2}, we briefly summarize the main ingredients
of the BLMSSM, then present the mass squared
matrices for neutral scalar sectors and the mass matrices for exotic quarks, respectively. We discuss the
corrections on the mass squared matrix of CP-even Higgs from exotic fields in section \ref{sec3},
and present the decay widths for $h^0\rightarrow\gamma\gamma,\;VV^*\;(V=Z,\;W)$ in section \ref{sec4}.
The numerical analyses are given in section \ref{sec5}, and our conclusions are summarized in section \ref{sec6}.

\section{A supersymmtric extension of the SM where B and L are local gauge symmetries\label{sec2}}
\indent\indent
When B and L are local gauge symmetries, one can enlarge the local gauge group of the SM to
$SU(3)_{_C}\otimes SU(2)_{_L}\otimes U(1)_{_Y}\otimes U(1)_{_B}\otimes U(1)_{_L}$.
In the supersymmetric extension of the SM proposed in Ref.\cite{BL_h,BL_h1}, the exotic superfields
include the new quarks $\hat{Q}_{_4}\sim(3,\;2,\;1/6,\;B_{_4},\;0)$,
$\hat{U}_{_4}^c\sim(\bar{3},\;1,\;-2/3,\;-B_{_4},\;0)$,
$\hat{D}_{_4}^c\sim(\bar{3},\;1,\;1/3,\;-B_{_4},\;0)$,
$\hat{Q}_{_5}^c\sim(\bar{3},\;2,\;-1/6,\;-(1+B_{_4}),\;0)$, $\hat{U}_{_5}\sim(3,\;1,\;2/3,\;1+B_{_4},\;0)$,
$\hat{D}_{_5}\sim(3,\;1,\;-1/3,\;1+B_{_4},\;0)$,
and the new leptons $\hat{L}_{_4}\sim(1,\;2,\;-1/2,\;0,\;L_{_4})$,
$\hat{E}_{_4}^c\sim(1,\;1,\;1,\;0,\;-L_{_4})$, $\hat{N}_{_4}^c\sim(1,\;1,\;0,\;0,\;-L_{_4})$,
$\hat{L}_{_5}^c\sim(1,\;2,\;1/2,\;0,\;-(3+L_{_4}))$, $\hat{E}_{_5}\sim(1,\;1,\;-1,\;0,\;3+L_{_4})$,
$\hat{N}_{_5}\sim(1,\;1,\;0,\;0,\;3+L_{_4})$ to cancel the $B$ and $L$ anomalies.
The 'brand new' Higgs superfields $\hat{\Phi}_{_B}\sim(1,\;1,\;0,\;1,\;0)$ and
$\hat{\varphi}_{_B}\sim(1,\;1,\;0,\;-1,\;0)$ acquire nonzero vacuum expectation values (VEVs) to break
Baryon number spontaneously. Meanwhile, nonzero VEVs of $\Phi_{_B}$ and $\phi_{_B}$ also
induce the large masses for exotic quarks.
In addition, the superfields $\hat{\Phi}_{_L}\sim(1,\;1,\;0,\;0,\;-2)$ and
$\hat{\varphi}_{_L}\sim(1,\;1,\;0,\;0,\;2)$ acquire nonzero VEVs to break Lepton number spontaneously.
In order to avoid stability for the exotic quarks, the model also includes the 
superfields $\hat{X}\sim(1,\;1,\;0,\;2/3+B_{_4},\;0)$,
$\hat{X}^\prime\sim(1,\;1,\;0,\;-(2/3+B_{_4}),\;0)$. Actually, the lightest one can be a dark matter candidate.
The superpotential of the model is written as
\begin{eqnarray}
&&{\cal W}_{_{BLMSSM}}={\cal W}_{_{MSSM}}+{\cal W}_{_B}+{\cal W}_{_L}+{\cal W}_{_X}\;,
\label{superpotential1}
\end{eqnarray}
where ${\cal W}_{_{MSSM}}$ is superpotential of the MSSM, and
\begin{eqnarray}
&&{\cal W}_{_B}=\lambda_{_Q}\hat{Q}_{_4}\hat{Q}_{_5}^c\hat{\Phi}_{_B}+\lambda_{_U}\hat{U}_{_4}^c\hat{U}_{_5}
\hat{\varphi}_{_B}+\lambda_{_D}\hat{D}_{_4}^c\hat{D}_{_5}\hat{\varphi}_{_B}+\mu_{_B}\hat{\Phi}_{_B}\hat{\varphi}_{_B}
\nonumber\\
&&\hspace{1.2cm}
+Y_{_{u_4}}\hat{Q}_{_4}\hat{H}_{_u}\hat{U}_{_4}^c+Y_{_{d_4}}\hat{Q}_{_4}\hat{H}_{_d}\hat{D}_{_4}^c
+Y_{_{u_5}}\hat{Q}_{_5}^c\hat{H}_{_d}\hat{U}_{_5}+Y_{_{d_5}}\hat{Q}_{_5}^c\hat{H}_{_u}\hat{D}_{_5}\;,
\nonumber\\
%%%%%%%%%%%%%%%%%%%%%%%%%%%%%%%%%%%%%%%%%%%%%%%%%%%%%%%%%%%%%
&&{\cal W}_{_L}=Y_{_{e_4}}\hat{L}_{_4}\hat{H}_{_d}\hat{E}_{_4}^c+Y_{_{\nu_4}}\hat{L}_{_4}\hat{H}_{_u}\hat{N}_{_4}^c
+Y_{_{e_5}}\hat{L}_{_5}^c\hat{H}_{_u}\hat{E}_{_5}+Y_{_{\nu_5}}\hat{L}_{_5}^c\hat{H}_{_d}\hat{N}_{_5}
\nonumber\\
&&\hspace{1.2cm}
+Y_{_\nu}\hat{L}\hat{H}_{_u}\hat{N}^c+\lambda_{_{N^c}}\hat{N}^c\hat{N}^c\hat{\varphi}_{_L}
+\mu_{_L}\hat{\Phi}_{_L}\hat{\varphi}_{_L}\;,
\nonumber\\
%%%%%%%%%%%%%%%%%%%%%%%%%%%%%%%%%%%%%%%%%%%%%%%%%%%%%%%%%%%%%
&&{\cal W}_{_X}=\lambda_1\hat{Q}\hat{Q}_{_5}^c\hat{X}+\lambda_2\hat{U}^c\hat{U}_{_5}\hat{X}^\prime
+\lambda_3\hat{D}^c\hat{D}_{_5}\hat{X}^\prime+\mu_{_X}\hat{X}\hat{X}^\prime\;.
\label{superpotential-BL}
\end{eqnarray}
In the superpotential above, the exotic quarks obtain ${\rm TeV}$ scale masses after $\Phi_{_B},\;\varphi_{_B}$
acquire nonzero VEVs, and the nonzero VEV of $\varphi_{_L}$ implements the seesaw mechanism for the tiny
neutrino masses. Correspondingly, the soft breaking terms are generally given as
\begin{eqnarray}
&&{\cal L}_{_{soft}}={\cal L}_{_{soft}}^{MSSM}-(m_{_{\tilde{N}^c}}^2)_{_{IJ}}\tilde{N}_I^{c*}\tilde{N}_J^c
-m_{_{\tilde{Q}_4}}^2\tilde{Q}_{_4}^\dagger\tilde{Q}_{_4}-m_{_{\tilde{U}_4}}^2\tilde{U}_{_4}^{c*}\tilde{U}_{_4}^c
-m_{_{\tilde{D}_4}}^2\tilde{D}_{_4}^{c*}\tilde{D}_{_4}^c
\nonumber\\
&&\hspace{1.3cm}
-m_{_{\tilde{Q}_5}}^2\tilde{Q}_{_5}^{c\dagger}\tilde{Q}_{_5}^c-m_{_{\tilde{U}_5}}^2\tilde{U}_{_5}^*\tilde{U}_{_5}
-m_{_{\tilde{D}_5}}^2\tilde{D}_{_5}^*\tilde{D}_{_5}-m_{_{\tilde{L}_4}}^2\tilde{L}_{_4}^\dagger\tilde{L}_{_4}
-m_{_{\tilde{\nu}_4}}^2\tilde{\nu}_{_4}^{c*}\tilde{\nu}_{_4}^c
\nonumber\\
&&\hspace{1.3cm}
-m_{_{\tilde{E}_4}}^2\tilde{e}_{_4}^{c*}\tilde{e}_{_4}^c-m_{_{\tilde{L}_5}}^2\tilde{L}_{_5}^{c\dagger}\tilde{L}_{_5}^c
-m_{_{\tilde{\nu}_5}}^2\tilde{\nu}_{_5}^*\tilde{\nu}_{_5}-m_{_{\tilde{E}_5}}^2\tilde{e}_{_5}^*\tilde{e}_{_5}
-m_{_{\Phi_{_B}}}^2\Phi_{_B}^*\Phi_{_B}
\nonumber\\
&&\hspace{1.3cm}
-m_{_{\varphi_{_B}}}^2\varphi_{_B}^*\varphi_{_B}-m_{_{\Phi_{_L}}}^2\Phi_{_L}^*\Phi_{_L}
-m_{_{\varphi_{_L}}}^2\varphi_{_L}^*\varphi_{_L}-\Big(m_{_B}\lambda_{_B}\lambda_{_B}
+m_{_L}\lambda_{_L}\lambda_{_L}+h.c.\Big)
\nonumber\\
&&\hspace{1.3cm}
+\Big\{A_{_{u_4}}Y_{_{u_4}}\tilde{Q}_{_4}H_{_u}\tilde{U}_{_4}^c+A_{_{d_4}}Y_{_{d_4}}\tilde{Q}_{_4}H_{_d}\tilde{D}_{_4}^c
+A_{_{u_5}}Y_{_{u_5}}\tilde{Q}_{_5}^cH_{_d}\tilde{U}_{_5}+A_{_{d_5}}Y_{_{d_5}}\tilde{Q}_{_5}^cH_{_u}\tilde{D}_{_5}
\nonumber\\
&&\hspace{1.3cm}
+A_{_{BQ}}\lambda_{_Q}\tilde{Q}_{_4}\tilde{Q}_{_5}^c\Phi_{_B}+A_{_{BU}}\lambda_{_U}\tilde{U}_{_4}^c\tilde{U}_{_5}\varphi_{_B}
+A_{_{BD}}\lambda_{_D}\tilde{D}_{_4}^c\tilde{D}_{_5}\varphi_{_B}+B_{_B}\mu_{_B}\Phi_{_B}\varphi_{_B}
+h.c.\Big\}
\nonumber\\
&&\hspace{1.3cm}
+\Big\{A_{_{e_4}}Y_{_{e_4}}\tilde{L}_{_4}H_{_d}\tilde{E}_{_4}^c+A_{_{N_4}}Y_{_{N_4}}\tilde{L}_{_4}H_{_u}\tilde{N}_{_4}^c
+A_{_{e_5}}Y_{_{e_5}}\tilde{L}_{_5}^cH_{_u}\tilde{E}_{_5}+A_{_{N_5}}Y_{_{\nu_5}}\tilde{L}_{_5}^cH_{_d}\tilde{N}_{_5}
\nonumber\\
&&\hspace{1.3cm}
+A_{_N}Y_{_N}\tilde{L}H_{_u}\tilde{N}^c+A_{_{N^c}}\lambda_{_{N^c}}\tilde{N}^c\tilde{N}^c\varphi_{_L}
+B_{_L}\mu_{_L}\Phi_{_L}\varphi_{_L}+h.c.\Big\}
\nonumber\\
&&\hspace{1.3cm}
+\Big\{A_1\lambda_1\tilde{Q}\tilde{Q}_{_5}^cX+A_2\lambda_2\tilde{U}^c\tilde{U}_{_5}X^\prime
+A_3\lambda_3\tilde{D}^c\tilde{D}_{_5}X^\prime+B_{_X}\mu_{_X}XX^\prime+h.c.\Big\}\;,
\label{soft-breaking}
\end{eqnarray}
where ${\cal L}_{_{soft}}^{MSSM}$ is soft breaking terms of the MSSM, $\lambda_B,\;\lambda_L$
are gauginos of $U(1)_{_B}$ and $U(1)_{_L}$, respectively.
After the $SU(2)_L$ doublets $H_{_u},\;H_{_d}$ and $SU(2)_L$ singlets $\Phi_{_B},\;\varphi_{_B},\;\Phi_{_L},\;
\varphi_{_L}$ acquire the nonzero VEVs $\upsilon_{_u},\;\upsilon_{_d},\;\upsilon_{_{B}},\;\overline{\upsilon}_{_{B}}$,
and $\upsilon_{_L},\;\overline{\upsilon}_{_L}$,
\begin{eqnarray}
&&H_{_u}=\left(\begin{array}{c}H_{_u}^+\\{1\over\sqrt{2}}\Big(\upsilon_{_u}+H_{_u}^0+iP_{_u}^0\Big)\end{array}\right)\;,
\nonumber\\
&&H_{_d}=\left(\begin{array}{c}{1\over\sqrt{2}}\Big(\upsilon_{_d}+H_{_d}^0+iP_{_d}^0\Big)\\H_{_d}^-\end{array}\right)\;,
\nonumber\\
&&\Phi_{_B}={1\over\sqrt{2}}\Big(\upsilon_{_B}+\Phi_{_B}^0+iP_{_B}^0\Big)\;,
\nonumber\\
&&\varphi_{_B}={1\over\sqrt{2}}\Big(\overline{\upsilon}_{_B}+\varphi_{_B}^0+i\overline{P}_{_B}^0\Big)\;,
\nonumber\\
&&\Phi_{_L}={1\over\sqrt{2}}\Big(\upsilon_{_L}+\Phi_{_L}^0+iP_{_L}^0\Big)\;,
\nonumber\\
&&\varphi_{_L}={1\over\sqrt{2}}\Big(\overline{\upsilon}_{_L}+\varphi_{_L}^0+i\overline{P}_{_L}^0\Big)\;,
\label{VEVs}
\end{eqnarray}
the local gauge symmetry $SU(2)_{_L}\otimes U(1)_{_Y}\otimes U(1)_{_B}\otimes U(1)_{_L}$ is broken
down to the electromagnetic symmetry $U(1)_{_e}$, where
\begin{eqnarray}
&&G^\pm=\cos\beta H_{_d}^\pm+\sin\beta H_{_u}^\pm
\label{Goldstone1}
\end{eqnarray}
denotes the charged Goldstone boson, and
\begin{eqnarray}
&&G^0=\cos\beta P_{_d}^0+\sin\beta P_{_u}^0\;,\nonumber\\
&&G_{_B}^0=\cos\beta_{_B} P_{_B}^0+\sin\beta_{_B}\overline{P}_{_B}^0\;,\nonumber\\
&&G_{_L}^0=\cos\beta_{_L} P_{_L}^0+\sin\beta_{_L}\overline{P}_{_L}^0
\label{Goldstone2}
\end{eqnarray}
denote the neutral Goldstone bosons, respectively. Here $\tan\beta=\upsilon_{_u}/\upsilon_{_d},\;
\tan\beta_{_B}=\overline{\upsilon}_{_B}/\upsilon_{_B}$, and $\tan\beta_{_L}=\overline{\upsilon}_{_L}/\upsilon_{_L}$.
Correspondingly, the physical neutral pseudoscalar fields are
\begin{eqnarray}
&&A^0=-\sin\beta P_{_d}^0+\cos\beta P_{_u}^0\;,\nonumber\\
&&A_{_B}^0=-\sin\beta_{_B} P_{_B}^0+\cos\beta_{_B}\overline{P}_{_B}^0\;,\nonumber\\
&&A_{_L}^0=-\sin\beta_{_L} P_{_L}^0+\cos\beta_{_L}\overline{P}_{_L}^0\;.
\label{neutral-pseudoscalar}
\end{eqnarray}
At tree level, the masses for those particles are respectively formulated as
\begin{eqnarray}
&&m_{_{A^0}}^2={B\mu\over\cos\beta\sin\beta}\;,\nonumber\\
&&m_{_{A_{_B}^0}}^2={B_{_B}\mu_{_B}\over\cos\beta_{_B}\sin\beta_{_B}}\;,\nonumber\\
&&m_{_{A_{_L}^0}}^2={B_{_L}\mu_{_L}\over\cos\beta_{_L}\sin\beta_{_L}}\;.
\label{pseudoscalar-mass}
\end{eqnarray}
Meanwhile the charged Higgs is
\begin{eqnarray}
&&H^\pm=-\sin\beta H_{_d}^\pm+\cos\beta H_{_u}^\pm
\label{charged-Higgs}
\end{eqnarray}
with the tree level mass squared
\begin{eqnarray}
&&m_{_{H^\pm}}^2=m_{_{A^0}}^2+m_{_{\rm W}}^2\;.
\label{charged-mass}
\end{eqnarray}
In the two Higgs doublet sector, the mass squared matrix of neutral CP-even Higgs is diagonalized by the rotation
\begin{eqnarray}
&&\left(\begin{array}{l}H^0\\h^0\end{array}\right)=\left(\begin{array}{cc}\cos\alpha&\sin\alpha\\
-\sin\alpha&\cos\alpha\end{array}\right)\left(\begin{array}{l}H_{_d}^0\\H_{_u}^0\end{array}\right)\;,
\label{CP-even-Higgs}
\end{eqnarray}
where $h^0$ is the lightest neutral CP-even Higgs.

In the basis $(\Phi_{_B}^0,\;\varphi_{_B}^0)$, the mass squared matrix is
\begin{eqnarray}
&&{\cal M}_{_{EB}}^2=\left(\begin{array}{ll}m_{_{Z_B}}^2\cos^2\beta_{_B}+m_{_{A_{_B}^0}}^2\sin^2\beta_{_B},\;&
(m_{_{Z_B}}^2+m_{_{A_{_B}^0}}^2)\cos\beta_{_B}\sin\beta_{_B}\\
(m_{_{Z_B}}^2+m_{_{A_{_B}^0}}^2)\cos\beta_{_B}\sin\beta_{_B},\;&
m_{_{Z_B}}^2\sin^2\beta_{_B}+m_{_{A_{_B}^0}}^2\cos^2\beta_{_B}
\end{array}\right)\;,
\label{CPevenB-mass}
\end{eqnarray}
where $m_{_{Z_B}}^2=g_{_B}^2(\upsilon_{_B}^2+\overline{\upsilon}_{_B}^2)$ is mass squared of
the neutral $U(1)_{_B}$ gauge boson $Z_{_B}$. Defining the mixing angle $\alpha_{_B}$ through
\begin{eqnarray}
&&\tan2\alpha_{_B}={m_{_{Z_B}}^2+m_{_{A_{_B}^0}}^2\over m_{_{Z_B}}^2-m_{_{A_{_B}^0}}^2}
\tan2\beta_{_B}\;,
\label{Mixing-B}
\end{eqnarray}
we obtain two mass eigenstates as
\begin{eqnarray}
&&\left(\begin{array}{l}H_{_B}^0\\ h_{_B}^0\end{array}\right)=
\left(\begin{array}{cc}\cos\alpha_{_B}&\sin\alpha_{_B}\\-\sin\alpha_{_B}&\cos\alpha_{_B}\end{array}\right)
\left(\begin{array}{l}\Phi_{_B}^0\\ \varphi_{_B}^0\end{array}\right)\;.
\label{Mass-eigenstates-B}
\end{eqnarray}

Similarly the mass squared matrix for $(\Phi_{_L}^0,\;\varphi_{_L}^0)$
is written as
\begin{eqnarray}
&&{\cal M}_{_{EL}}^2=\left(\begin{array}{ll}m_{_{Z_L}}^2\cos^2\beta_{_L}+m_{_{A_{_L}^0}}^2\sin^2\beta_{_L},\;&
(m_{_{Z_L}}^2+m_{_{A_{_L}^0}}^2)\cos\beta_{_L}\sin\beta_{_L}\\
(m_{_{Z_L}}^2+m_{_{A_{_L}^0}}^2)\cos\beta_{_L}\sin\beta_{_L},\;&
m_{_{Z_L}}^2\sin^2\beta_{_L}+m_{_{A_{_L}^0}}^2\cos^2\beta_{_L}
\end{array}\right)\;,
\label{CPevenL-mass}
\end{eqnarray}
with $m_{_{Z_L}}^2=4g_{_L}^2(\upsilon_{_L}^2+\overline{\upsilon}_{_L}^2)$ denoting mass squared of
the neutral $U(1)_{_L}$ gauge boson $Z_{_L}$.

In four-component Dirac spinors, the mass matrix for exotic quarks with charged $2/3$ is
\begin{eqnarray}
&&-{\cal L}_{_{t^\prime}}^{mass}=\left(\begin{array}{ll}\bar{t}_{_{4R}}^\prime,&\bar{t}_{_{5R}}^\prime\end{array}\right)
\left(\begin{array}{ll}{1\over\sqrt{2}}\lambda_{_Q}\upsilon_{_B},&-{1\over\sqrt{2}}Y_{_{u_5}}\upsilon_{_d}\\
-{1\over\sqrt{2}}Y_{_{u_4}}\upsilon_{_u},&{1\over\sqrt{2}}\lambda_{_u}\overline{\upsilon}_{_B}
\end{array}\right)\left(\begin{array}{l}t_{_{4L}}^\prime\\t_{_{5L}}^\prime\end{array}\right)+h.c.
\label{Qmass-matrix-2/3}
\end{eqnarray}
Using the unitary transformations
\begin{eqnarray}
&&\left(\begin{array}{l}t_{_{4L}}^\prime\\t_{_{5L}}^\prime\end{array}\right)
=U_{_{t^\prime}}^\dagger\cdot\left(\begin{array}{l}t_{_{4L}}\\t_{_{5L}}\end{array}\right)\;,\;\;
\left(\begin{array}{l}t_{_{4R}}^\prime\\t_{_{5R}}^\prime\end{array}\right)
=W_{_{t^\prime}}^\dagger\cdot\left(\begin{array}{l}t_{_{4R}}\\t_{_{5R}}\end{array}\right)\;,
\label{Qmixing-2/3-a}
\end{eqnarray}
we diagonalize the mass matrix for the vector quarks with charged $2/3$:
\begin{eqnarray}
&&W_{_{t^\prime}}^\dagger\cdot\left(\begin{array}{ll}{1\over\sqrt{2}}\lambda_{_Q}\upsilon_{_B},&
-{1\over\sqrt{2}}Y_{_{u_5}}\upsilon_{_d}\\-{1\over\sqrt{2}}Y_{_{u_4}}\upsilon_{_u},&
{1\over\sqrt{2}}\lambda_{_u}\overline{\upsilon}_{_B}\end{array}\right)\cdot U_{_{t^\prime}}
={\it diag}\Big(m_{_{t_4}},\;m_{_{t_5}}\Big)
\label{Qmixing-2/3-b}
\end{eqnarray}
Similarly we can write the mass matrix for the exotic quarks with charged $-1/3$ as
\begin{eqnarray}
&&-{\cal L}_{_{b^\prime}}^{mass}=\left(\begin{array}{ll}\bar{b}_{_{4R}},&\bar{b}_{_{5R}}\end{array}\right)
\left(\begin{array}{ll}-{1\over\sqrt{2}}\lambda_{_Q}\upsilon_{_B},&-{1\over\sqrt{2}}Y_{_{d_5}}\upsilon_{_u}\\
-{1\over\sqrt{2}}Y_{_{d_4}}\upsilon_{_d},&{1\over\sqrt{2}}\lambda_{_d}\overline{\upsilon}_{_B}
\end{array}\right)\left(\begin{array}{l}b_{_{4L}}\\b_{_{5L}}\end{array}\right)+h.c.
\label{Qmass-matrix-1/3}
\end{eqnarray}
Adopting the unitary transformations
\begin{eqnarray}
&&\left(\begin{array}{l}b_{_{4L}}^\prime\\b_{_{5L}}^\prime\end{array}\right)
=U_{_{b^\prime}}^\dagger\cdot\left(\begin{array}{l}b_{_{4L}}\\b_{_{5L}}\end{array}\right)\;,\;\;
\left(\begin{array}{l}b_{_{4R}}^\prime\\b_{_{5R}}^\prime\end{array}\right)
=W_{_{b^\prime}}^\dagger\cdot\left(\begin{array}{l}b_{_{4R}}\\b_{_{5R}}\end{array}\right)\;,
\label{Qmixing-1/3-a}
\end{eqnarray}
one can diagonalize mass matrix for the vector quarks with charged $-1/3$ as
\begin{eqnarray}
&&W_{_{b^\prime}}^\dagger\cdot\left(\begin{array}{ll}-{1\over\sqrt{2}}\lambda_{_Q}\upsilon_{_B},&
-{1\over\sqrt{2}}Y_{_{d_5}}\upsilon_{_u}\\-{1\over\sqrt{2}}Y_{_{d_4}}\upsilon_{_d},&
{1\over\sqrt{2}}\lambda_{_d}\overline{\upsilon}_{_B}\end{array}\right)\cdot U_{_{b^\prime}}
={\it diag}\Big(m_{_{b_4}},\;m_{_{b_5}}\Big)\;.
\label{Qmixing-1/3-b}
\end{eqnarray}
Assuming CP conservation in exotic quark sector, we then derive the flavor conservative
couplings between the lightest neutral CP-even Higgs and charged $2/3$ exotic quarks:
\begin{eqnarray}
&&{\cal L}_{_{Ht^\prime t^\prime}}={1\over\sqrt{2}}\sum\limits_{i=1}^2
\Big[Y_{_{u_4}}(W_{_t}^T)_{_{i2}}(U_{_t})_{_{1i}}\cos\alpha
+Y_{_{u_5}}(W_{_t}^T)_{_{i1}}(U_{_t})_{_{2i}}\sin\alpha\Big]
h^0\overline{t}_{_{i+3}}t_{_{i+3}}\;,
\label{Higgs-EQ-2/3}
\end{eqnarray}
where $T$ represents the transposing transformation of a matrix. In a similar way,  the flavor
conservative couplings between the lightest neutral CP-even Higgs and charged $-1/3$ exotic quarks
are written as
\begin{eqnarray}
&&{\cal L}_{_{Hb^\prime b^\prime}}={1\over\sqrt{2}}\sum\limits_{i=1}^2
\Big[Y_{_{d_4}}(W_{_b}^T)_{_{i2}}(U_{_b})_{_{1i}}\sin\alpha
-Y_{_{d_5}}(W_{_b}^T)_{_{i1}}(U_{_b})_{_{2i}}\cos\alpha\Big]
h^0\overline{b}_{_{i+3}}b_{_{i+3}}\;.
\label{Higgs-EQ-1/3}
\end{eqnarray}

Using the superpotential in Eq.(\ref{superpotential1}) and the soft breaking terms,
we write the mass squared matrices for exotic scalar quarks as
\begin{eqnarray}
&&-{\cal L}_{_{\widetilde{EQ}}}^{mass}=\tilde{t}^{\prime\dagger}\cdot
{\cal M}_{\tilde{t}^\prime}^2\cdot\tilde{t}^\prime
+\tilde{b}^{\prime\dagger}\cdot {\cal M}_{\tilde{b}^\prime}^2\cdot\tilde{b}^\prime
\label{SQmass-2/3}
\end{eqnarray}
with $\tilde{t}^{\prime T}=(\tilde{Q}_{_4}^{1},\;\tilde{U}_{_4}^{c*},\;\tilde{Q}_{_5}^{2c*},\;
\tilde{U}_{_5})$, $\tilde{b}^{\prime T}=(\tilde{Q}_{_4}^{2},\;\tilde{D}_{_4}^{c*},
\;\tilde{Q}_{_5}^{1c*},\;\tilde{D}_{_5}^{*})$. The concrete expressions for $4\times4$
mass squared matrices ${\cal M}_{\tilde{t}^\prime}^2,\;{\cal M}_{\tilde{b}^\prime}^2$
are given in appendix \ref{app2}, and the couplings between the lightest neutral CP-even Higgs
and exotic scalar quarks are collected in appendix \ref{app3}.

\section{The lightest CP-even Higgs mass\label{sec3}}
\indent\indent
It is well known since quite some time that radiative corrections modify the tree level mass squared
matrix of neutral Higgs substantially in the MSSM, where the main effect originates from one-loop diagrams involving the top
quark and its scalar partners $\tilde{t}_{1,2}$ \cite{Haber1}. In order to obtain masses of the
neutral CP-even Higgs reasonably, we should include the radiative corrections from exotic fermions
and corresponding supersymmetric partners in the BLMSSM. Then, the mass squared matrix for the
neutral CP-even Higgs in the basis $(H_d^0,\;H_u^0)$ is written as
\begin{eqnarray}
&&{\cal M}^2_{even}=\left(\begin{array}{ll}M_{11}^2+\Delta_{11}&M_{12}^2+\Delta_{12}\\
M_{12}^2+\Delta_{12}&M_{22}^2+\Delta_{22}\end{array}\right)\;,
\label{M-CPE}
\end{eqnarray}
where
\begin{eqnarray}
&&M_{11}^2=m_{_{\rm Z}}^2\cos^2\beta+m_{_{A^0}}^2\sin^2\beta\;,
\nonumber\\
&&M_{12}^2=-(m_{_{\rm Z}}^2+m_{_{A^0}}^2)\sin\beta\cos\beta\;,
\nonumber\\
&&M_{22}^2=m_{_{\rm Z}}^2\sin^2\beta+m_{_{A^0}}^2\cos^2\beta\;,
\nonumber\\
\label{M-CPE1}
\end{eqnarray}
and $m_{_{A^0}}$ denotes the pseudo-scalar Higgs mass at tree level.
The radiative corrections originate from the MSSM sector, exotic fermions and corresponding scalar fermions
respectively in this model:
\begin{eqnarray}
&&\Delta_{11}=\Delta_{11}^{MSSM}+\Delta_{11}^{B}+\Delta_{11}^{L}\;,
\nonumber\\
&&\Delta_{12}=\Delta_{12}^{MSSM}+\Delta_{12}^{B}+\Delta_{12}^{L}\;,
\nonumber\\
&&\Delta_{22}=\Delta_{22}^{MSSM}+\Delta_{22}^{B}+\Delta_{22}^{L}\;.
\nonumber\\
\label{M-CPE2}
\end{eqnarray}
Here the concrete expressions for $\Delta_{11}^{MSSM}$, $\Delta_{12}^{MSSM}$, $\Delta_{22}^{MSSM}$
at two-loop level can be found in literature\cite{2loop-HiggsM}, the one-loop radiative corrections from
exotic quark fields are formulated as\cite{1loop-HiggsM}
\begin{eqnarray}
%%%%%%%%%%%%%%%%%%%%%%%%%%%%%%%%%%%%%%%%%%%%%%%%%%%%%%%%%%
&&\Delta_{11}^{B}={3G_{_F}Y_{_{u_4}}^4\upsilon^4\over4\sqrt{2}\pi^2\sin^2\beta}\cdot
{\mu^2(A_{_{u_4}}-\mu\cot\beta)^2\over(m_{_{\tilde{t}_1^\prime}}^2-m_{_{\tilde{t}_2^\prime}}^2)^2}
g(m_{_{\tilde{t}_1^\prime}},m_{_{\tilde{t}_2^\prime}})
\nonumber\\
&&\hspace{1.2cm}
+{3G_{_F}Y_{_{u_5}}^4\upsilon^4\over4\sqrt{2}\pi^2\cos^2\beta}\Big\{\ln{m_{_{\tilde{t}_3^\prime}}m_{_{\tilde{t}_4^\prime}}
\over m_{_{t_5}}^2}+{A_{_{u_5}}(A_{_{u_5}}-\mu\tan\beta)\over m_{_{\tilde{t}_3^\prime}}^2-m_{_{\tilde{t}_4^\prime}}^2}
\ln{m_{_{\tilde{t}_3^\prime}}^2\over m_{_{\tilde{t}_4^\prime}}^2}
\nonumber\\
&&\hspace{1.2cm}
+{A_{_{u_5}}^2(A_{_{u_5}}-\mu\tan\beta)^2\over(m_{_{\tilde{t}_3^\prime}}^2-m_{_{\tilde{t}_4^\prime}}^2)^2}
g(m_{_{\tilde{t}_3^\prime}},m_{_{\tilde{t}_4^\prime}})\Big\}
\nonumber\\
&&\hspace{1.2cm}
+{3G_{_F}Y_{_{d_4}}^4\upsilon^4\over4\sqrt{2}\pi^2\cos^2\beta}\Big\{\ln{m_{_{\tilde{b}_1^\prime}}m_{_{\tilde{b}_2^\prime}}
\over m_{_{b_4}}^2}+{A_{_{d_4}}(A_{_{d_4}}-\mu\tan\beta)\over m_{_{\tilde{b}_1^\prime}}^2-m_{_{\tilde{b}_2^\prime}}^2}
\ln{m_{_{\tilde{b}_1^\prime}}^2\over m_{_{\tilde{b}_2^\prime}}^2}
\nonumber\\
&&\hspace{1.2cm}
+{A_{_{d_4}}^2(A_{_{d_4}}-\mu\tan\beta)^2\over(m_{_{\tilde{b}_1^\prime}}^2-m_{_{\tilde{b}_2^\prime}}^2)^2}
g(m_{_{\tilde{b}_1^\prime}},m_{_{\tilde{b}_2^\prime}})\Big\}
\nonumber\\
&&\hspace{1.2cm}
+{3G_{_F}Y_{_{d_5}}^4\upsilon^4\over4\sqrt{2}\pi^2\sin^2\beta}\cdot
{\mu^2(A_{_{d_5}}-\mu\cot\beta)^2\over(m_{_{\tilde{b}_3^\prime}}^2-m_{_{\tilde{b}_4^\prime}}^2)^2}
g(m_{_{\tilde{b}_3^\prime}},m_{_{\tilde{b}_4^\prime}})
\;,\nonumber\\
%%%%%%%%%%%%%%%%%%%%%%%%%%%%%%%%%%%%%%%%%%%%%%%%%%%%%%%%%%
&&\Delta_{12}^{B}={3G_{_F}Y_{_{u_4}}^4\upsilon^4\over8\sqrt{2}\pi^2\sin^2\beta}\cdot
{\mu(-A_{_{u_4}}+\mu\cot\beta)\over m_{_{\tilde{t}_1^\prime}}^2-m_{_{\tilde{t}_2^\prime}}^2}
\Big\{\ln{m_{_{\tilde{t}_1^\prime}}\over m_{_{\tilde{t}_2^\prime}}}+{A_{_{u_4}}(A_{_{u_4}}-\mu\cot\beta)
\over m_{_{\tilde{t}_1^\prime}}^2-m_{_{\tilde{t}_2^\prime}}^2}g(m_{_{\tilde{t}_1^\prime}},m_{_{\tilde{t}_2^\prime}})\Big\}
\nonumber\\
&&\hspace{1.2cm}
+{3G_{_F}Y_{_{u_5}}^4\upsilon^4\over8\sqrt{2}\pi^2\cos^2\beta}\cdot
{\mu(-A_{_{u_5}}+\mu\tan\beta)\over m_{_{\tilde{t}_3^\prime}}^2-m_{_{\tilde{t}_4^\prime}}^2}
\Big\{\ln{m_{_{\tilde{t}_3^\prime}}\over m_{_{\tilde{t}_4^\prime}}}+{A_{_{u_5}}(A_{_{u_5}}-\mu\tan\beta)
\over m_{_{\tilde{t}_3^\prime}}^2-m_{_{\tilde{t}_4^\prime}}^2}g(m_{_{\tilde{t}_3^\prime}},m_{_{\tilde{t}_4^\prime}})\Big\}
\nonumber\\
&&\hspace{1.2cm}
+{3G_{_F}Y_{_{d_4}}^4\upsilon^4\over8\sqrt{2}\pi^2\cos^2\beta}\cdot
{\mu(-A_{_{d_4}}+\mu\tan\beta)\over m_{_{\tilde{d}_1^\prime}}^2-m_{_{\tilde{d}_2^\prime}}^2}
\Big\{\ln{m_{_{\tilde{d}_1^\prime}}\over m_{_{\tilde{d}_2^\prime}}}+{A_{_{d_4}}(A_{_{d_4}}-\mu\tan\beta)
\over m_{_{\tilde{d}_1^\prime}}^2-m_{_{\tilde{d}_2^\prime}}^2}g(m_{_{\tilde{d}_1^\prime}},m_{_{\tilde{d}_2^\prime}})\Big\}
\nonumber\\
&&\hspace{1.2cm}
+{3G_{_F}Y_{_{d_5}}^4\upsilon^4\over8\sqrt{2}\pi^2\sin^2\beta}\cdot
{\mu(-A_{_{d_5}}+\mu\cot\beta)\over m_{_{\tilde{b}_3^\prime}}^2-m_{_{\tilde{b}_4^\prime}}^2}
\Big\{\ln{m_{_{\tilde{b}_3^\prime}}\over m_{_{\tilde{b}_4^\prime}}}+{A_{_{d_5}}(A_{_{d_5}}-\mu\cot\beta)
\over m_{_{\tilde{b}_3^\prime}}^2-m_{_{\tilde{b}_4^\prime}}^2}g(m_{_{\tilde{b}_3^\prime}},m_{_{\tilde{b}_4^\prime}})\Big\}
\;,\nonumber\\
%%%%%%%%%%%%%%%%%%%%%%%%%%%%%%%%%%%%%%%%%%%%%%%%%%%%%%%%%%
&&\Delta_{22}^{B}={3G_{_F}Y_{_{u_4}}^4\upsilon^4\over4\sqrt{2}\pi^2\sin^2\beta}\Big\{\ln{m_{_{\tilde{t}_1^\prime}}m_{_{\tilde{t}_2^\prime}}
\over m_{_{t_4}}^2}+{A_{_{u_4}}(A_{_{u_4}}-\mu\cot\beta)\over m_{_{\tilde{t}_1^\prime}}^2-m_{_{\tilde{t}_2^\prime}}^2}
\ln{m_{_{\tilde{t}_1^\prime}}^2\over m_{_{\tilde{t}_2^\prime}}^2}
\nonumber\\
&&\hspace{1.2cm}
+{A_{_{u_4}}^2(A_{_{u_4}}-\mu\cot\beta)^2\over(m_{_{\tilde{t}_1^\prime}}^2-m_{_{\tilde{t}_2^\prime}}^2)^2}
g(m_{_{\tilde{t}_1^\prime}},m_{_{\tilde{t}_2^\prime}})\Big\}
\nonumber\\
&&\hspace{1.2cm}
+{3G_{_F}Y_{_{u_5}}^4\upsilon^4\over4\sqrt{2}\pi^2\cos^2\beta}\cdot
{\mu^2(A_{_{u_5}}-\mu\tan\beta)^2\over(m_{_{\tilde{t}_3^\prime}}^2-m_{_{\tilde{t}_4^\prime}}^2)^2}
g(m_{_{\tilde{t}_3^\prime}},m_{_{\tilde{t}_4^\prime}})
\nonumber\\
&&\hspace{1.2cm}
+{3G_{_F}Y_{_{d_4}}^4\upsilon^4\over4\sqrt{2}\pi^2\cos^2\beta}\cdot
{\mu^2(A_{_{d_4}}-\mu\tan\beta)^2\over(m_{_{\tilde{b}_1^\prime}}^2-m_{_{\tilde{b}_2^\prime}}^2)^2}
g(m_{_{\tilde{b}_1^\prime}},m_{_{\tilde{b}_2^\prime}})
\nonumber\\
&&\hspace{1.2cm}
+{3G_{_F}Y_{_{d_5}}^4\upsilon^4\over4\sqrt{2}\pi^2\sin^2\beta}\Big\{\ln{m_{_{\tilde{b}_3^\prime}}m_{_{\tilde{b}_4^\prime}}
\over m_{_{b_5}}^2}+{A_{_{d_5}}(A_{_{d_5}}-\mu\cot\beta)\over m_{_{\tilde{b}_3^\prime}}^2-m_{_{\tilde{b}_4^\prime}}^2}
\ln{m_{_{\tilde{b}_3^\prime}}^2\over m_{_{\tilde{b}_4^\prime}}^2}
\nonumber\\
&&\hspace{1.2cm}
+{A_{_{d_5}}^2(A_{_{d_5}}-\mu\cot\beta)^2\over(m_{_{\tilde{b}_3^\prime}}^2-m_{_{\tilde{b}_4^\prime}}^2)^2}
g(m_{_{\tilde{b}_3^\prime}},m_{_{\tilde{b}_4^\prime}})\Big\}\;,
\label{M-CPE3}
\end{eqnarray}
here $\upsilon=\sqrt{\upsilon_{_u}^2+\upsilon_{_d}^2}\simeq246\;{\rm GeV}$ and
\begin{eqnarray}
&&g(x,y)=1-{x^2+y^2\over x^2-y^2}\ln{x\over y}\;.
\label{M-CPE4}
\end{eqnarray}
To derive the results presented in Eq.(\ref{M-CPE3}), we adopt the assumption $|\lambda_{_Q}\upsilon_{_B}|,\;
|\lambda_{_u}\overline{\upsilon}_{_B}|,\;|\lambda_{_d}\overline{\upsilon}_{_B}|\gg |Y_{_{u_4}}\upsilon|,\;
|Y_{_{u_5}}\upsilon|,\;|Y_{_{d_4}}\upsilon|,\;|Y_{_{d_5}}\upsilon|$ in our calculation.
Similarly, one can obtain the one-loop radiative corrections from exotic lepton fields presented
in appendix \ref{app4}.

One most stringent constraint on parameter space of the BLMSSM is that the mass squared
matrix in Eq.(\ref{M-CPE}) should produce an eigenvalue around $(125\;{\rm GeV})^2$
as mass squared of the lightest neutral CP-even Higgs.
The current combination of the ATLAS and CMS data gives\cite{CMS, ATLAS,CMS-ATLAS}:
\begin{eqnarray}
&&m_{_{h^0}}=125.9\pm2.1\;{\rm GeV}\;,
\label{M-h0}
\end{eqnarray}
this fact constrains parameter space of the BLMSSM strongly.

\section{$gg\rightarrow h^0$ and $h^0\rightarrow\gamma\gamma,\;ZZ^*,\;WW^*$\label{sec4}}
\indent\indent
The Higgs is produced chiefly through the gluon fusion at the LHC.  In the SM, the leading order (LO) contributions originate
from the one loop diagram which involves virtual top quarks. The cross section for this process is known to
the next-to-next-to-leading order (NNLO)\cite{NNLO} which can enhance the LO result by 80-100\%. Furthermore, any new particle
which strongly couples with the Higgs can significantly modified this cross section. 
In supersymmetric extension of the SM considered here,
the LO decay width for the process $h^0\rightarrow gg$ is given by (see Ref.\cite{Gamma1} and references therein)
\begin{eqnarray}
&&\Gamma_{_{NP}}(h^0\rightarrow gg)={G_{_F}\alpha_s^2m_{_{h^0}}^3\over64\sqrt{2}\pi^3}
\Big|\sum\limits_qg_{_{h^0qq}}A_{1/2}(x_q)
+\sum\limits_{\tilde q}g_{_{h^0\tilde{q}\tilde{q}}}{m_{_{\rm Z}}^2\over m_{_{\tilde q}}^2}A_{0}(x_{_{\tilde{q}}})\Big|^2\;,
\label{hgg}
\end{eqnarray}
with $x_a=m_{_{h^0}}^2/(4m_a^2)$. In addition, $q=t,\;b,\;t_{_4},\;t_{_5},\;b_{_4},\;b_{_5}$
and $\tilde{q}=\tilde{t}_{_{1,2}},\;\tilde{b}_{_{1,2}},\;\tilde{\cal U}_i,\;\tilde{\cal D}_i\;(i=1,\;2,\;3,\;4)$.
The concrete expressions for $g_{_{h^0tt}},\;g_{_{h^0bb}},\;g_{_{h^0\tilde{t}_i\tilde{t}_i}}
,\;g_{_{h^0\tilde{b}_i\tilde{b}_i}},\;(i=1,\;2)$ can be found in literature\cite{BL_h1}, and
\begin{eqnarray}
&&g_{_{h^0t_4t_4}}=-{\sqrt{2}m_{_{\rm W}}s_{_{\rm W}}\over em_{_{t_4}}}
\Big[Y_{_{u_4}}(W_{_t}^T)_{_{12}}(U_{_t})_{_{11}}\cos\alpha
+Y_{_{u_5}}(W_{_t}^T)_{_{11}}(U_{_t})_{_{21}}\sin\alpha\Big]
\;,\nonumber\\
&&g_{_{h^0t_5t_5}}=-{\sqrt{2}m_{_{\rm W}}s_{_{\rm W}}\over em_{_{t_5}}}
\Big[Y_{_{u_4}}(W_{_t}^T)_{_{22}}(U_{_t})_{_{12}}\cos\alpha
+Y_{_{u_5}}(W_{_t}^T)_{_{21}}(U_{_t})_{_{22}}\sin\alpha\Big]
\;,\nonumber\\
&&g_{_{h^0b_4b_4}}=-{\sqrt{2}m_{_{\rm W}}s_{_{\rm W}}\over em_{_{b_4}}}
\Big[Y_{_{d_4}}(W_{_b}^T)_{_{12}}(U_{_b})_{_{11}}\sin\alpha
-Y_{_{d_5}}(W_{_b}^T)_{_{11}}(U_{_b})_{_{21}}\cos\alpha\Big]
\;,\nonumber\\
&&g_{_{h^0b_5b_5}}=-{\sqrt{2}m_{_{\rm W}}s_{_{\rm W}}\over em_{_{b_5}}}
\Big[Y_{_{d_4}}(W_{_b}^T)_{_{22}}(U_{_b})_{_{12}}\sin\alpha
-Y_{_{d_5}}(W_{_b}^T)_{_{21}}(U_{_b})_{_{22}}\cos\alpha\Big]
\;,\nonumber\\
&&g_{_{h^0\tilde{\cal U}_i\tilde{\cal U}_i}}=-{m_{_{\rm W}}^2s_{_{\rm W}}\over em_{_{\tilde{\cal U}_i}}^2}
\Big[\xi_{_{uii}}^S\cos\alpha-\xi_{_{dii}}^S\sin\alpha\Big]\;,\;\;(i=1,\;2,\;3,\;4)
\;,\nonumber\\
&&g_{_{h^0\tilde{\cal D}_i\tilde{\cal D}_i}}=-{m_{_{\rm W}}^2s_{_{\rm W}}\over em_{_{\tilde{\cal D}_i}}^2}
\Big[\eta_{_{uii}}^S\cos\alpha-\eta_{_{dii}}^S\sin\alpha\Big]\;,\;\;(i=1,\;2,\;3,\;4)\;.
\label{g-coupling1}
\end{eqnarray}
Here, we adopt the abbreviation $s_{_{\rm W}}=\sin\theta_{_{\rm W}}$ with $\theta_{_{\rm W}}$
denoting the Weinberg angle. Furthermore, $e$ is the electromagnetic coupling constant, and
the concrete expressions of $\xi_{_{uii}}^S,\;\xi_{_{dii}}^S,\;\eta_{_{uii}}^S,\;\eta_{_{dii}}^S$ can be
found in appendix \ref{app3}.  The form factors $A_{1/2},\;A_0$ in Eq.(\ref{hgg}) are defined as
\begin{eqnarray}
&&A_{1/2}(x)=2\Big[x+(x-1)g(x)\Big]/x^2\;,\nonumber\\
&&A_0(x)=-(x-g(x))/x^2\;,
\label{loop-function1}
\end{eqnarray}
with
\begin{eqnarray}
&&g(x)=\left\{\begin{array}{l}\arcsin^2\sqrt{x},\;x\le1\\
-{1\over4}\Big[\ln{1+\sqrt{1-1/x}\over1-\sqrt{1-1/x}}-i\pi\Big]^2,\;x>1\;.\end{array}\right.
\label{g-function}
\end{eqnarray}

The Higgs to diphoton decay is also obtained from loop diagrams, the LO contributions are derived from the
one loop diagrams containing virtual charged gauge boson $W^\pm$ or virtual top quarks
in the SM. In the BLMSSM, the exotic fermions $t_{_{4,5}},\;b_{_{4,5}},\;e_{_{4,5}}$
together with their supersymmetric partners contribute the corrections to the decay width
of Higgs to diphoton at LO, the corresponding expression is written as
\begin{eqnarray}
&&\Gamma_{_{NP}}(h^0\rightarrow\gamma\gamma)={G_{_F}\alpha^2m_{_{h^0}}^3\over128\sqrt{2}\pi^3}
\Big|\sum\limits_fN_cQ_{_f}^2g_{_{h^0ff}}A_{1/2}(x_f)+g_{_{h^0WW}}A_1(x_{_{\rm W}})
\nonumber\\
&&\hspace{3.2cm}
+g_{_{h^0H^+H^-}}{m_{_{\rm W}}^2\over m_{_{H^\pm}}^2}A_0(x_{_{H^\pm}})
+\sum\limits_{i=1}^2g_{_{h^0\chi_i^+\chi_i^-}}{m_{_{\rm W}}\over m_{_{\chi_i}}}A_{1/2}(x_{_{\chi_i}})
\nonumber\\
&&\hspace{3.2cm}
+\sum\limits_{\tilde f}N_cQ_{_f}^2g_{_{h^0\tilde{f}\tilde{f}}}{m_{_{\rm Z}}^2\over m_{_{\tilde f}}^2}
A_{0}(x_{_{\tilde{f}}})\Big|^2\;,
\label{hpp}
\end{eqnarray}
where $g_{_{h^0WW}}=\sin(\beta-\alpha)$, the concrete expression for the loop functions $A_1$ is
\begin{eqnarray}
&&A_1(x)=-\Big[2x^2+3x+3(2x-1)g(x)\Big]/x^2\;.\nonumber\\
\label{loop-function2}
\end{eqnarray}
The concrete expressions for $g_{_{h^0\chi_i^+\chi_i^-}},\;g_{_{h^0H^+H^-}}$
and the couplings between the lightest neutral CP-even Higgs and exotic leptons/sleptons can also
be found in literature\cite{BL_h1}.

For the lightest neutral CP-even Higgs around $125\;{\rm GeV}$ mass, it can decay through the modes
$h^0\rightarrow ZZ^*,\;h^0\rightarrow WW^*$ where $Z^*/W^*$ denotes the off-shell neutral/charged
electroweak gauge bosons. Summing over all channels available to the $W^*$ or $Z^*$, one can write
the widths as\cite{Keung1,HtoVV-SUSY}
\begin{eqnarray}
&&\Gamma(h^0\rightarrow WW^*)={3e^4m_{_{h^0}}\over512\pi^3s_{_{\rm W}}^4}|g_{_{h^0WW}}|^2
F({m_{_{\rm W}}\over m_{_{h^0}}}),\;\nonumber\\
&&\Gamma(h^0\rightarrow ZZ^*)={e^4m_{_{h^0}}\over2048\pi^3s_{_{\rm W}}^4c_{_{\rm W}}^4}|g_{_{h^0ZZ}}|^2
\Big(7-{40\over3}s_{_{\rm W}}^2+{160\over9}s_{_{\rm W}}^4\Big)F({m_{_{\rm Z}}\over m_{_{h^0}}}),\;\nonumber\\
\label{h-WW*ZZ*}
\end{eqnarray}
with $g_{_{h^0ZZ}}=g_{_{h^0WW}}$ and the abbreviation $c_{_{\rm W}}=\cos\theta_{_{\rm W}}$.
The form factor $F(x)$ is given as
\begin{eqnarray}
&&F(x)=-(1-x^2)\Big({47\over2}x^2-{13\over2}+{1\over x^2}\Big)-3(1-6x^2+4x^4)\ln x
\nonumber\\
&&\hspace{1.5cm}
+{3(1-8x^2+20x^4)\over\sqrt{4x^2-1}}\cos^{-1}\Big({3x^2-1\over2x^3}\Big)\;.\nonumber\\
\label{form-factor1}
\end{eqnarray}

Besides the Higgs discovery the ATLAS and CMS experiments have both observed an excess
in Higgs production and decay into diphoton channel which is a factor $1.4\sim2$ times larger than
the SM expectations. The observed signals for the diphoton and $ZZ^*,\;WW^*$ channels are 
quantified by the ratios
\begin{eqnarray}
&&R_{\gamma\gamma}={\Gamma_{_{NP}}(h_0\rightarrow gg)\Gamma_{_{NP}}(h_0\rightarrow\gamma\gamma)\over
\Gamma_{_{SM}}(h_0\rightarrow gg)\Gamma_{_{SM}}(h_0\rightarrow\gamma\gamma)}
\;,\nonumber\\
&&R_{VV^*}={\Gamma_{_{NP}}(h_0\rightarrow gg)\Gamma_{_{NP}}(h_0\rightarrow VV^*)\over
\Gamma_{_{SM}}(h_0\rightarrow gg)\Gamma_{_{SM}}(h_0\rightarrow VV^*)}
\;,\;\;(V=Z,\;W)\;.\nonumber\\
\label{signal}
\end{eqnarray}
The current values of the ratios are \cite{CMS,ATLAS,CMS-ATLAS}:
\begin{eqnarray}
&&{\rm ATLAS+CMS}:\;\;R_{\gamma\gamma}=1.77\pm0.33\;,
\nonumber\\
&&{\rm ATLAS+CMS}:\;\;R_{VV^*}=0.94\pm0.40\;,(V=Z,\;W)\;.
\label{signal-exp}
\end{eqnarray}
Note that the combinations of the ATLAS and CMS results are taken from Ref.\cite{CMS-ATLAS}.

\section{Numerical analyses\label{sec5}}
\indent\indent

%%%%%%%%%%%%%%%%%%%%%%%%%%%%%%%%%%%%%%%%%%%%%%%%%%%%%
\begin{figure}[h]
\setlength{\unitlength}{1mm}
\centering
\includegraphics[width=4.0in]{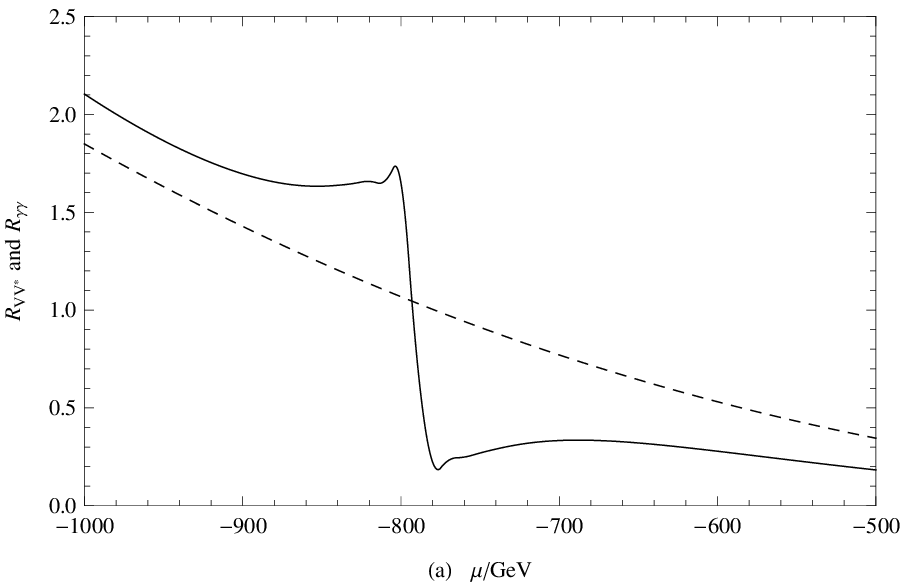}
\vspace{0.5cm}
\includegraphics[width=4.0in]{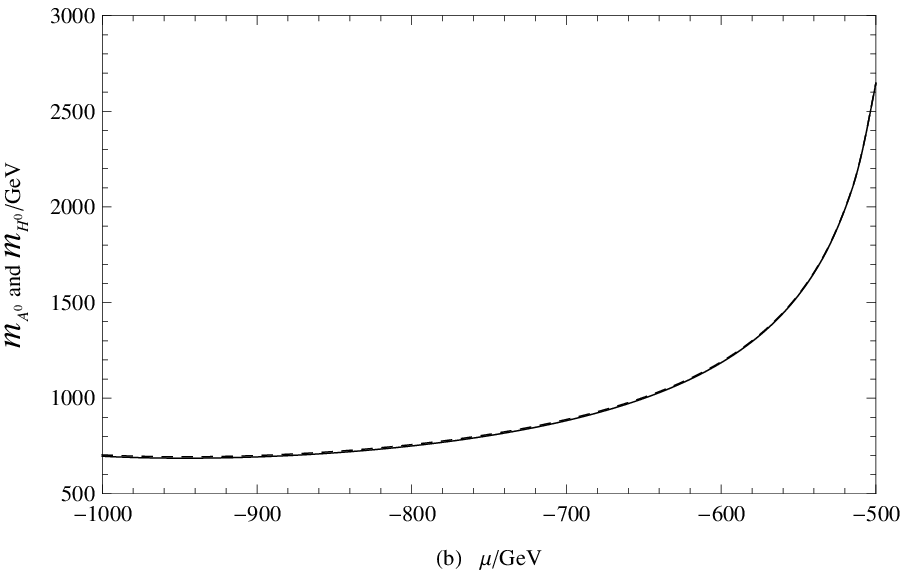}
\vspace{0cm}
\caption[]{As $Y_{_{u_5}}=0.7Y_b,\;Y_{_{d_5}}=0.13Y_t,\;m_{_{\tilde{Q}_4}}=790\;{\rm GeV}$
and $A_t=-1\;{\rm TeV}$, (a) $R_{\gamma\gamma}$ (solid-line) and $R_{VV^*}$ (dashed-line) vary with the
parameter $\mu$, and (b)$m_{_{A^0}}$ (solid-line) and $m_{_{H^0}}$ (dashed-line) vary with the parameter
$\mu$, respectively.}
\label{fig1}
\end{figure}
%%%%%%%%%%%%%%%%%%%%%%%%%%%%%%%%%%%%%%%%%%%%%%%%%%%%%

As mentioned above, the most stringent constraint on the parameter space is that the
$2\times2$ mass squared matrix in Eq.(\ref{M-CPE}) should predict the lightest eigenvector
with a mass $m_{_{h_0}}\simeq125.9\;{\rm GeV}$. In order to obtain the final results coinciding with this condition,
we require the tree level mass of CP-odd Higgs $m_{_{A^0}}$ satisfying
\begin{eqnarray}
&&m_{_{A^0}}^2={m_{_{h_0}}^2(m_{_{\rm z}}^2-m_{_{h_0}}^2+\Delta_{_{11}}+\Delta_{_{22}})-m_{_{\rm z}}^2
\Delta_{_A}+\Delta_{_{12}}^2-\Delta_{_{11}}\Delta_{_{22}}\over -m_{_{h_0}}^2+m_{_{\rm z}}^2\cos^22\beta
+\Delta_{_B}}\;,
\label{Higgs-mass1}
\end{eqnarray}
where
\begin{eqnarray}
&&\Delta_{_A}=\sin^2\beta\Delta_{_{11}}+\cos^2\beta\Delta_{_{22}}+\sin2\beta \Delta_{_{12}}
\;,\nonumber\\
&&\Delta_{_B}=\cos^2\beta\Delta_{_{11}}+\sin^2\beta\Delta_{_{22}}+\sin2\beta \Delta_{_{12}}\;.
\label{Higgs-mass2}
\end{eqnarray}
%%%%%%%%%%%%%%%%%%%%%%%%%%%%%%%%%%%%%%%%%%%%%%%%%%%%%
\begin{figure}[h]
\setlength{\unitlength}{1mm}
\centering
\includegraphics[width=4.0in]{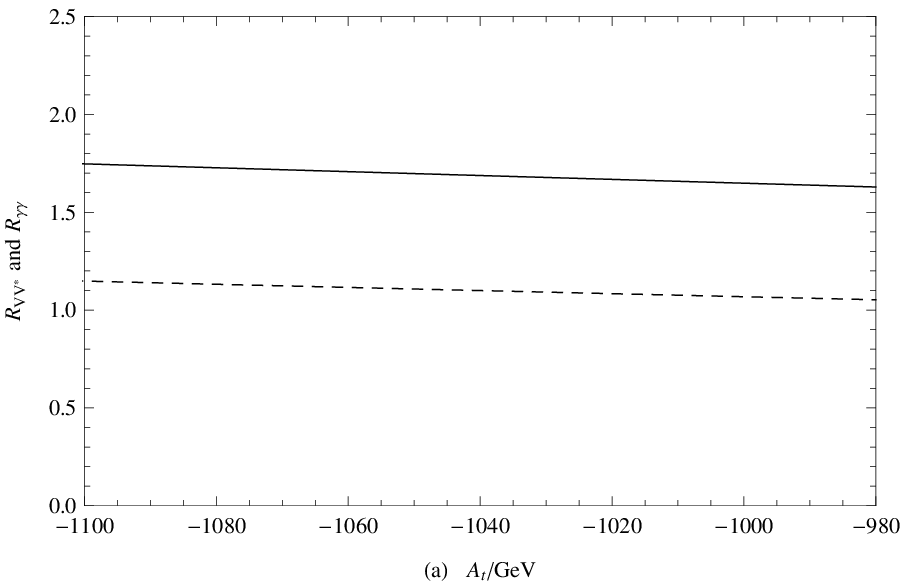}
\vspace{0.5cm}
\includegraphics[width=4.0in]{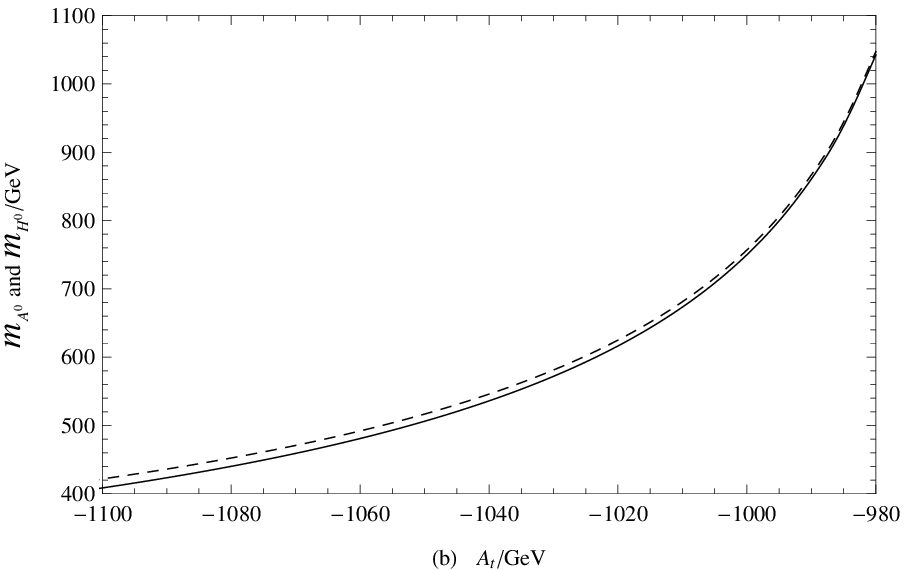}
\vspace{0cm}
\caption[]{As $Y_{_{u_5}}=0.7Y_b,\;Y_{_{d_5}}=0.13Y_t,\;m_{_{\tilde{Q}_4}}=790\;{\rm GeV}$
and $\mu=-800\;{\rm GeV}$, (a) $R_{\gamma\gamma}$ (solid-line) and $R_{VV^*}$ (dashed-line) vary with the
parameter $A_t$, and (b)$m_{_{A^0}}$ (solid-line) and $m_{_{H^0}}$ (dashed-line) vary with the parameter
$A_t$, respectively.}
\label{fig2}
\end{figure}
%%%%%%%%%%%%%%%%%%%%%%%%%%%%%%%%%%%%%%%%%%%%%%%%%%%%%

Through scanning the parameter space, we find the evaluations on $R_{\gamma\gamma}$,
$R_{VV^*}$ and masses of the heaviest CP-even Higgs and CP-odd Higgs depending on
$\tan\beta$ acutely as $m_{_{h^0}}=125.9\;{\rm GeV}$.
In our numerical analysis, we adopt the ansatz on relevant parameter space
\begin{eqnarray}
&&B_4=L_4={3\over2}\;,
\nonumber\\
&&m_{_{\tilde{Q}_3}}=m_{_{\tilde{U}_3}}=m_{_{\tilde{D}_3}}=1\;{\rm TeV}\;,
\nonumber\\
&&m_{_{\tilde{U}_4}}=m_{_{\tilde{D}_4}}=m_{_{\tilde{Q}_5}}=m_{_{\tilde{U}_5}}
=m_{_{\tilde{D}_5}}=1\;{\rm TeV}\;,\nonumber\\
&&m_{_{\tilde{L}_4}}=m_{_{\tilde{\nu}_4}}=m_{_{\tilde{E}_4}}=m_{_{\tilde{L}_5}}=m_{_{\tilde{\nu}_5}}
=m_{_{\tilde{E}_5}}=1\;{\rm TeV}\;,\nonumber\\
&&m_{_{Z_B}}=m_{_{Z_L}}=1\;{\rm TeV}\;,\nonumber\\
&&A_{_{\nu_4}}=A_{_{e_4}}=A_{_{\nu_5}}=A_{_{e_4}}=A_{_{d_4}}=A_{_{u_5}}=A_{_{d_5}}=550\;{\rm GeV}
\;,\nonumber\\
&&\upsilon_{_{B_t}}=\sqrt{\upsilon_{_B}^2+\overline{\upsilon}_{_B}^2}=3\;{\rm TeV}\;,\;\;
\upsilon_{_{L_t}}=\sqrt{\upsilon_{_L}^2+\overline{\upsilon}_{_L}^2}=3\;{\rm TeV}\;,
\nonumber\\
&&A_{_{BQ}}=A_{_{BU}}=A_{_{BD}}=-A_{_b}=1\;{\rm TeV}\;,
\nonumber\\
&&Y_{_{u_4}}=0.76\;Y_t\;,\;\;Y_{_{d_4}}=0.7\;Y_b\;,\;\;\lambda_{_Q}=\lambda_{_u}=\lambda_{_d}=0.5
\nonumber\\
&&m_2=750\;{\rm GeV}\;,\;\;\;\mu_{_B}=500\;{\rm GeV}\;,
\;,\nonumber\\
&&\tan\beta=\tan\beta_{_B}=\tan\beta_{_L}=2\;,\nonumber\\
\label{assumption1}
\end{eqnarray}
to reduce the number of free parameters in the model considered here.
Furthermore, we choose the masses for exotic leptons from Ref.\cite{BL_h1}:
\begin{eqnarray}
&&m_{_{\nu_4}}=m_{_{\nu_5}}=90\;{\rm GeV}\;,\;\;m_{_{e_4}}=m_{_{e_5}}=100\;{\rm GeV}\;.
\label{assumption2}
\end{eqnarray}

%%%%%%%%%%%%%%%%%%%%%%%%%%%%%%%%%%%%%%%%%%%%%%%%%%%%%
\begin{figure}[h]
\setlength{\unitlength}{1mm}
\centering
\includegraphics[width=4.0in]{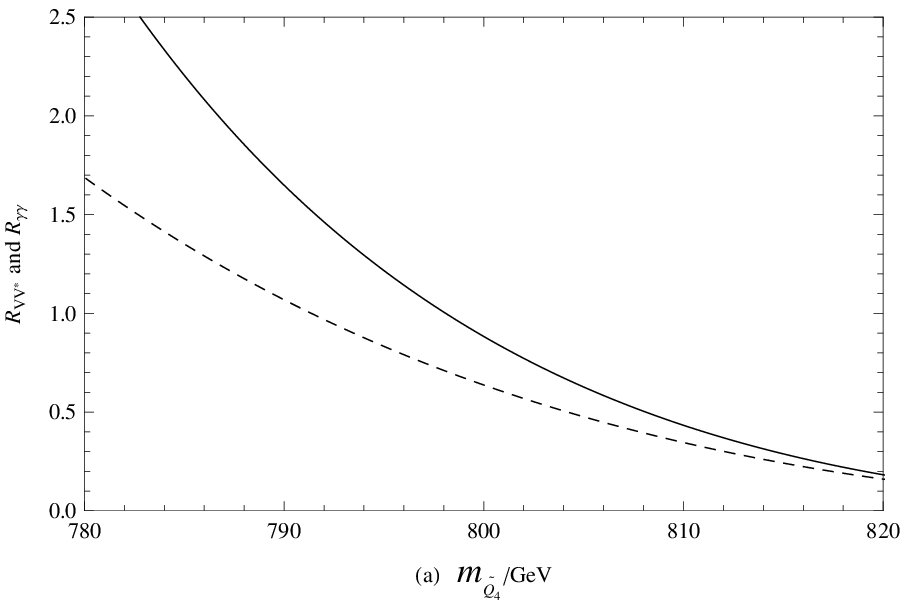}
\vspace{0.5cm}
\includegraphics[width=4.0in]{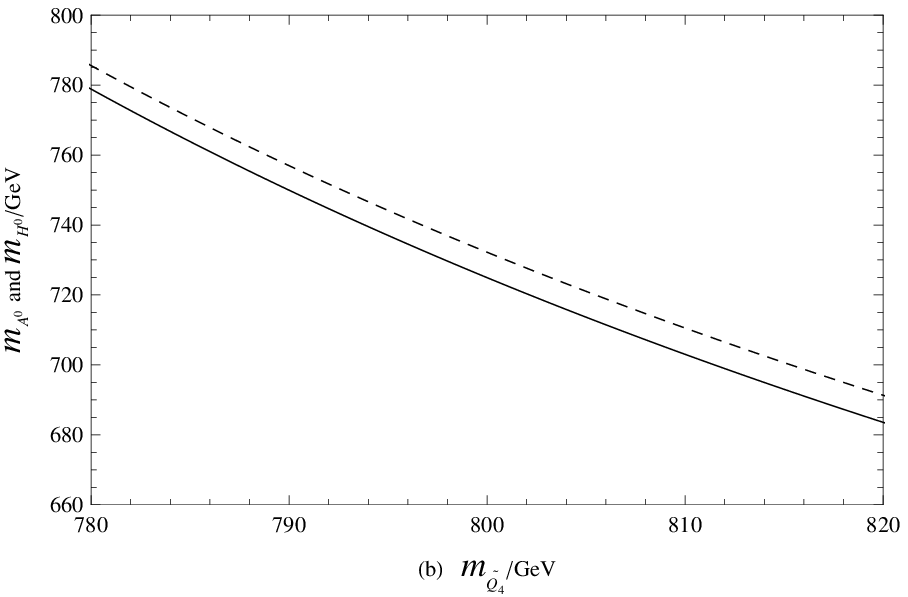}
\vspace{0cm}
\caption[]{As $Y_{_{u_5}}=0.7Y_b,\;Y_{_{d_5}}=0.13Y_t,\;A_t=-1\;{\rm TeV}$
and $\mu=-800\;{\rm GeV}$, (a) $R_{\gamma\gamma}$ (solid-line) and $R_{VV^*}$ (dashed-line) vary with the
parameter $m_{_{\tilde{Q}_4}}$, and (b)$m_{_{A^0}}$ (solid-line) and $m_{_{H^0}}$ (dashed-line) vary with the parameter
$m_{_{\tilde{Q}_4}}$, respectively.}
\label{fig3}
\end{figure}
%%%%%%%%%%%%%%%%%%%%%%%%%%%%%%%%%%%%%%%%%%%%%%%%%%%%%

For relevant parameters in the SM, we choose\cite{PDG}
\begin{eqnarray}
&&\alpha_s(m_{_{\rm Z}})=0.118\;,\;\;\alpha(m_{_{\rm Z}})=1/128\;,
\;\;s_{_{\rm W}}^2(m_{_{\rm Z}})=0.23\;,
\nonumber\\
&&m_t=174.2\;{\rm GeV}\;,\;\;m_b=4.2\;{\rm GeV}\;,\;\;
m_{_{\rm W}}=80.4\;{\rm GeV}\;.
\label{PDG-SM}
\end{eqnarray}

%%%%%%%%%%%%%%%%%%%%%%%%%%%%%%%%%%%%%%%%%%%%%%%%%%%%%
\begin{figure}[h]
\setlength{\unitlength}{1mm}
\centering
\includegraphics[width=4.0in]{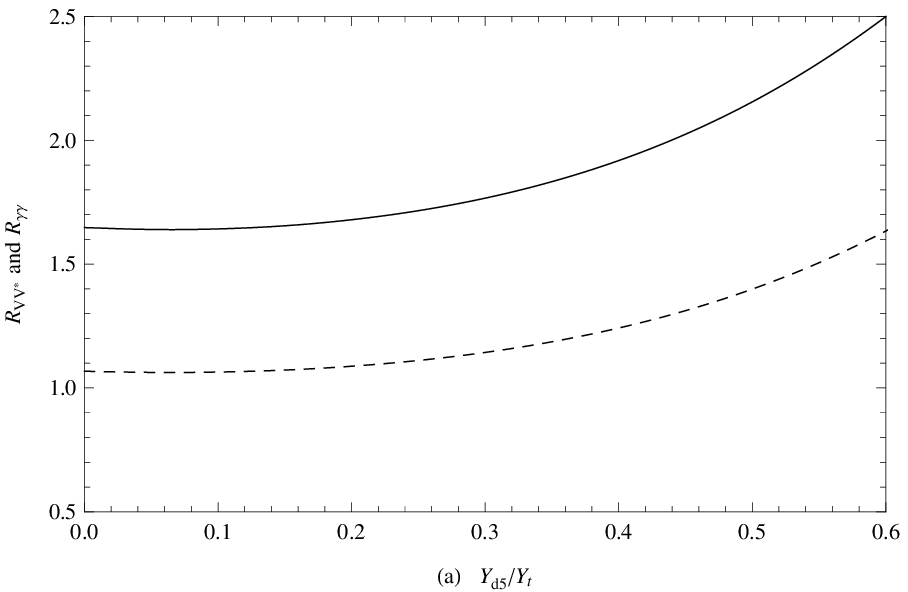}
\vspace{0.5cm}
\includegraphics[width=4.0in]{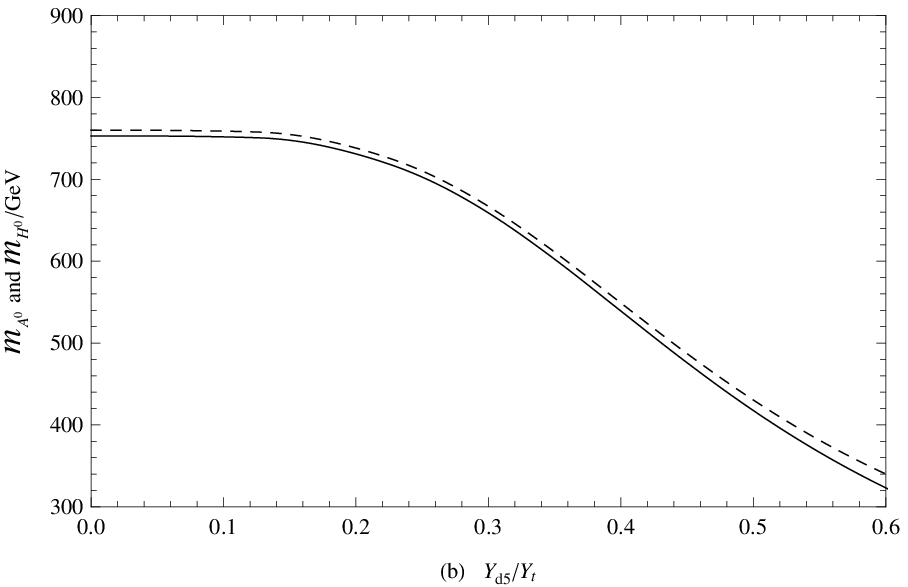}
\vspace{0cm}
\caption[]{As $Y_{_{u_5}}=0.7Y_b,\;A_t=-1\;{\rm TeV}$ and $\mu=-790\;{\rm GeV}$,
(a) $R_{\gamma\gamma}$ (solid-line) and $R_{VV^*}$ (dashed-line) vary with the
ratio $Y_{_{d_5}}/Y_t$, and (b)$m_{_{A^0}}$ (solid-line) and $m_{_{H^0}}$ (dashed-line) vary with
the ratio $Y_{_{d_5}}/Y_t$, respectively.}
\label{fig4}
\end{figure}
%%%%%%%%%%%%%%%%%%%%%%%%%%%%%%%%%%%%%%%%%%%%%%%%%%%%%

Considering that the CMS collaboration has excluded a SM Higgs with mass in the range $127.5\;{\rm GeV}-
600\;{\rm GeV}$, we require the theoretical evaluations on masses of the heaviest CP-even Higgs and CP-odd
Higgs respectively in the range $m_{_{A^0}}\ge600\;{\rm GeV},\;m_{_{H^0}}\ge600\;{\rm GeV}$. Choosing $Y_{_{u_5}}=
Y_{_{d_4}}=0.7Y_b,\;Y_{_{d_5}}=0.13Y_t,\;m_{_{\tilde{Q}_4}}=790\;{\rm GeV}$ and $A_t=-1\;{\rm TeV}$, we
plot $R_{\gamma\gamma}$ (solid-line) and $R_{VV^*}$ (dashed-line) varying with the parameter $\mu$
in Fig.\ref{fig1}(a), and plot $m_{_{A^0}}$ (solid-line) and $m_{_{H^0}}$ (dashed-line) varying with the parameter
$\mu$ in Fig.\ref{fig1}(b), respectively. Using our assumptions on relevant parameter space in the BLMSSM,
we find the theoretical evaluations on $R_{\gamma\gamma}$, $R_{VV^*}$, $m_{_{A^0}}$ and $m_{_{H^0}}$
depending on $\mu$ acutely. As $-900\;{\rm GeV}\le\mu\le-800\;{\rm GeV}$, the theoretical predictions
on $R_{\gamma\gamma}$ and $R_{VV^*}$ are all coincide with experimental data in Eq.(\ref{signal-exp}),
and masses of the heaviest CP-even Higgs and CP-odd Higgs $m_{_{A^0}}\sim m_{_{H^0}}\ge700\;{\rm GeV}$
simultaneously.

Another parameter  $A_t$ maybe affects the theoretical evaluations on $R_{\gamma\gamma}$ and $R_{VV^*}$
strongly here. Taking $Y_{_{u_5}}=Y_{_{d_4}}=0.7Y_b,\;Y_{_{d_5}}=0.13Y_t,\;m_{_{\tilde{Q}_4}}=790\;{\rm GeV}$
and $\mu=-800\;{\rm GeV}$, we depict $R_{\gamma\gamma}$ (solid-line) and $R_{VV^*}$ (dashed-line) varying
with the parameter $A_t$ in Fig.\ref{fig2}(a), and plot $m_{_{A^0}}$ (solid-line) and $m_{_{H^0}}$ (dashed-line)
varying with the parameter  $A_t$ in Fig.\ref{fig2}(b), respectively.  Under our assumptions on the parameter space,
the dependence of $R_{\gamma\gamma}$ and $R_{VV^*}$ on $A_t$ is very mild. Nevertheless, the theoretical
evaluations of $m_{_{A^0}}$ and $m_{_{H^0}}$ depend on $A_t$ strongly. When $A_t\ge-1\;{TeV}$, the theoretical
evaluations on $R_{\gamma\gamma}$ and $R_{VV^*}$ are all coincide with experimental data in Eq.(\ref{signal-exp}),
and masses of the heaviest CP-even Higgs and CP-odd Higgs $m_{_{A^0}}\sim m_{_{H^0}}\ge700\;{\rm GeV}$
meantime.

Besides those parameters existing in the MSSM already, the 'brand new' parameters in the BLMSSM
also affect the theoretical evaluations on $R_{\gamma\gamma}$, $R_{VV^*}$ and $m_{_{A^0}},\;m_{_{H^0}}$
strongly as $m_{_{h^0}}=125.9\;{\rm GeV}$.
In Fig.(\ref{fig3}), we investigate $R_{\gamma\gamma}$, $R_{VV^*}$ and $m_{_{A^0}},\;m_{_{H^0}}$
versus the soft mass of fourth generation left-handed scalar quarks $m_{_{\tilde{Q}_4}}$.
Where the solid line in Fig.(\ref{fig3})(a) represents $R_{\gamma\gamma}$ varying with $m_{_{\tilde{Q}_4}}$,
the dashed line in Fig.(\ref{fig3})(a) represents $R_{VV^*}$ varying with $m_{_{\tilde{Q}_4}}$,
the solid line in Fig.(\ref{fig3})(b) represents $m_{_{A^0}}$ varying with $m_{_{\tilde{Q}_4}}$,
the dashed line in Fig.(\ref{fig3})(b) represents $m_{_{H^0}}$ varying with $m_{_{\tilde{Q}_4}}$, respectively.
Actually, the theoretical evaluations on $R_{\gamma\gamma}$, $R_{VV^*}$, $m_{_{A^0}}$ and 
$m_{_{H^0}}$ decrease steeply with
the increasing of $m_{_{\tilde{Q}_4}}$. When $m_{_{\tilde{Q}_4}}\ge 800\;{\rm GeV}$, the theoretical
prediction on $R_{\gamma\gamma}$ already lies out the experimental range in Eq.(\ref{signal-exp}).
In Fig.(\ref{fig4}), we investigate the theoretical predictions on $R_{\gamma\gamma}$, $R_{VV^*}$ and $m_{_{A^0}},\;m_{_{H^0}}$
versus the Yukawa coupling of fifth generation down-type quark $Y_{_{d_5}}$.
Where the solid line in Fig.(\ref{fig4})(a) represents $R_{\gamma\gamma}$ varying with $Y_{_{d_5}}$,
the dashed line in Fig.(\ref{fig4})(a) represents $R_{VV^*}$ varying with $Y_{_{d_5}}$,
the solid line in Fig.(\ref{fig4})(b) represents $m_{_{A^0}}$ varying with $Y_{_{d_5}}$,
the dashed line in Fig.(\ref{fig4})(b) represents $m_{_{H^0}}$ varying with $Y_{_{d_5}}$, respectively.
In fact, the theoretical evaluations on $R_{\gamma\gamma}$ and $R_{VV^*}$ raise slowly with
increasing of the ratio $Y_{_{d_5}}/Y_t$. When $Y_{_{d_5}}/Y_t\ge 0.5$, the theoretical
predictions on $R_{\gamma\gamma},\;R_{VV^*}$ exceed the experimental range in Eq.(\ref{signal-exp}),
and the numerical evaluations on $m_{_{A^0}},\;m_{_{H^0}}$ are below $600\;{\rm GeV}$.

\section{Summary\label{sec6}}
\indent\indent
In framework of the BLMSSM, we attempt to account
for the experimental data on Higgs reported by ATLAS and CMS recently.
Assuming the Yukawa couplings between Higgs doublets and exotic quarks satisfying
$Y_{_{u_4}},\;Y_{_{d_5}}<Y_{_t}$ as well as $Y_{_{d_4}},\;Y_{_{u_5}}<Y_{_b}$, we find the theoretical
predictions on $R_{\gamma\gamma},\;R_{VV^*}$ fitting the experimental data in Eq.(\ref{signal-exp})
very well when $m_{_{h_0}}=125.9\;{\rm GeV}$.
Furthermore, the numerical evaluations on $m_{_{A^0}},\;m_{_{H^0}}$ exceed $600\;{\rm GeV}$
simultaneously in some parameter space of the BLMSSM.
\begin{acknowledgments}
\indent\indent
The work has been supported by the National Natural Science Foundation of China (NNSFC)
with Grant No. 11275036, No. 11047002 and Natural Science Fund of Hebei University
with Grant No. 2011JQ05, No. 2012-242.
\end{acknowledgments}

\appendix
\section{The couplings between neutral Higgs and exotic quarks\label{app1}}
\indent\indent
In the mass basis, the couplings between the neutral Higgs and charged $2/3$ exotic quarks
are written as
\begin{eqnarray}
&&{\cal L}_{_{Ht^\prime t^\prime}}={1\over\sqrt{2}}\sum\limits_{i,j=1}^2\Big\{
\Big[Y_{_{u_4}}(W_{_t}^\dagger)_{_{i2}}(U_{_t})_{_{1j}}\cos\alpha
+Y_{_{u_5}}(W_{_t}^\dagger)_{_{i1}}(U_{_t})_{_{2j}}\sin\alpha\Big]
h^0\overline{t}_{_{i+3}}P_{_L}t_{_{j+3}}
\nonumber\\
&&\hspace{1.8cm}
+\Big[Y_{_{u_4}}(U_{_t}^\dagger)_{_{i1}}(W_{_t})_{_{2j}}\cos\alpha
+Y_{_{u_5}}(U_{_t}^\dagger)_{_{i2}}(W_{_t})_{_{1j}}\sin\alpha\Big]
h^0\overline{t}_{_{i+3}}P_{_R}t_{_{j+3}}
\nonumber\\
&&\hspace{1.8cm}
+\Big[Y_{_{u_4}}(W_{_t}^\dagger)_{_{i2}}(U_{_t})_{_{1j}}\sin\alpha
-Y_{_{u_5}}(W_{_t}^\dagger)_{_{i1}}(U_{_t})_{_{2j}}\cos\alpha\Big]
H^0\overline{t}_{_{i+3}}P_{_L}t_{_{j+3}}
\nonumber\\
&&\hspace{1.8cm}
+\Big[Y_{_{u_4}}(U_{_t}^\dagger)_{_{i1}}(W_{_t})_{_{2j}}\sin\alpha
-Y_{_{u_5}}(U_{_t}^\dagger)_{_{i2}}(W_{_t})_{_{1j}}\cos\alpha\Big]
H^0\overline{t}_{_{i+3}}P_{_R}t_{_{j+3}}\Big\}
\nonumber\\
&&\hspace{1.8cm}
+{i\over\sqrt{2}}\sum\limits_{i,j=1}^2\Big\{
\Big[Y_{_{u_4}}(W_{_t}^\dagger)_{_{i2}}(U_{_t})_{_{1j}}\cos\beta
+Y_{_{u_5}}(W_{_t}^\dagger)_{_{i1}}(U_{_t})_{_{2j}}\sin\beta\Big]
A^0\overline{t}_{_{i+3}}P_{_L}t_{_{j+3}}
\nonumber\\
&&\hspace{1.8cm}
-\Big[Y_{_{u_4}}(U_{_t}^\dagger)_{_{i1}}(W_{_t})_{_{2j}}\cos\beta
+Y_{_{u_5}}(U_{_t}^\dagger)_{_{i2}}(W_{_t})_{_{1j}}\sin\beta\Big]
A^0\overline{t}_{_{i+3}}P_{_R}t_{_{j+3}}
\nonumber\\
&&\hspace{1.8cm}
+\Big[Y_{_{u_4}}(W_{_t}^\dagger)_{_{i2}}(U_{_t})_{_{1j}}\sin\beta
-Y_{_{u_5}}(W_{_t}^\dagger)_{_{i1}}(U_{_t})_{_{2j}}\cos\beta\Big]
G^0\overline{t}_{_{i+3}}P_{_L}t_{_{j+3}}
\nonumber\\
&&\hspace{1.8cm}
-\Big[Y_{_{u_4}}(U_{_t}^\dagger)_{_{i1}}(W_{_t})_{_{2j}}\sin\beta
-Y_{_{u_5}}(U_{_t}^\dagger)_{_{i2}}(W_{_t})_{_{1j}}\cos\beta\Big]
G^0\overline{t}_{_{i+3}}P_{_R}t_{_{j+3}}\Big\}
\nonumber\\
&&\hspace{1.8cm}
-{1\over\sqrt{2}}\sum\limits_{i,j=1}^2\Big\{
\Big[\lambda_{_u}(W_{_t}^\dagger)_{_{i2}}(U_{_t})_{_{2j}}\cos\alpha_{_B}
-\lambda_{_Q}(W_{_t}^\dagger)_{_{i1}}(U_{_t})_{_{1j}}\sin\alpha_{_B}\Big]
h_{_B}^0\overline{t}_{_{i+3}}P_{_L}t_{_{j+3}}
\nonumber\\
&&\hspace{1.8cm}
+\Big[\lambda_{_u}(U_{_t}^\dagger)_{_{i2}}(W_{_t})_{_{2j}}\cos\alpha_{_B}
-\lambda_{_Q}(U_{_t}^\dagger)_{_{i2}}(W_{_t})_{_{2j}}\sin\alpha_{_B}\Big]
h_{_B}^0\overline{t}_{_{i+3}}P_{_R}t_{_{j+3}}
\nonumber\\
&&\hspace{1.8cm}
+\Big[\lambda_{_u}(W_{_t}^\dagger)_{_{i2}}(U_{_t})_{_{2j}}\sin\alpha_{_B}
+\lambda_{_Q}(W_{_t}^\dagger)_{_{i1}}(U_{_t})_{_{1j}}\cos\alpha_{_B}\Big]
H_{_B}^0\overline{t}_{_{i+3}}P_{_L}t_{_{j+3}}
\nonumber\\
&&\hspace{1.8cm}
+\Big[\lambda_{_u}(U_{_t}^\dagger)_{_{i2}}(W_{_t})_{_{2j}}\sin\alpha_{_B}
+\lambda_{_Q}(U_{_t}^\dagger)_{_{i2}}(W_{_t})_{_{2j}}\cos\alpha_{_B}\Big]
H_{_B}^0\overline{t}_{_{i+3}}P_{_R}t_{_{j+3}}\Big\}
\nonumber\\
&&\hspace{1.8cm}
-{i\over\sqrt{2}}\sum\limits_{i,j=1}^2\Big\{
\Big[\lambda_{_u}(W_{_t}^\dagger)_{_{i2}}(U_{_t})_{_{2j}}\cos\beta_{_B}
-\lambda_{_Q}(W_{_t}^\dagger)_{_{i1}}(U_{_t})_{_{1j}}\sin\beta_{_B}\Big]
A_{_B}^0\overline{t}_{_{i+3}}P_{_L}t_{_{j+3}}
\nonumber\\
&&\hspace{1.8cm}
-\Big[\lambda_{_u}(U_{_t}^\dagger)_{_{i2}}(W_{_t})_{_{2j}}\cos\beta_{_B}
-\lambda_{_Q}(U_{_t}^\dagger)_{_{i2}}(W_{_t})_{_{2j}}\sin\beta_{_B}\Big]
A_{_B}^0\overline{t}_{_{i+3}}P_{_R}t_{_{j+3}}
\nonumber\\
&&\hspace{1.8cm}
+\Big[\lambda_{_u}(W_{_t}^\dagger)_{_{i2}}(U_{_t})_{_{2j}}\sin\beta_{_B}
+\lambda_{_Q}(W_{_t}^\dagger)_{_{i1}}(U_{_t})_{_{1j}}\cos\beta_{_B}\Big]
G_{_B}^0\overline{t}_{_{i+3}}P_{_L}t_{_{j+3}}
\nonumber\\
&&\hspace{1.8cm}
-\Big[\lambda_{_u}(U_{_t}^\dagger)_{_{i2}}(W_{_t})_{_{2j}}\sin\beta_{_B}
+\lambda_{_Q}(U_{_t}^\dagger)_{_{i2}}(W_{_t})_{_{2j}}\cos\beta_{_B}\Big]
H_{_B}^0\overline{t}_{_{i+3}}P_{_R}t_{_{j+3}}\Big\}
\label{app1-2/3}
\end{eqnarray}
Similarly, the couplings between the neutral Higgs and charged $-1/3$ exotic quarks
are written as
\begin{eqnarray}
&&{\cal L}_{_{Hb^\prime b^\prime}}={1\over\sqrt{2}}\sum\limits_{i,j=1}^2\Big\{
\Big[Y_{_{d_4}}(W_{_b}^\dagger)_{_{i2}}(U_{_b})_{_{1j}}\sin\alpha
-Y_{_{d_5}}(W_{_b}^\dagger)_{_{i1}}(U_{_b})_{_{2j}}\cos\alpha\Big]
h^0\overline{b}_{_{i+3}}P_{_L}b_{_{j+3}}
\nonumber\\
&&\hspace{1.8cm}
+\Big[Y_{_{d_4}}(U_{_b}^\dagger)_{_{i1}}(W_{_b})_{_{2j}}\sin\alpha
-Y_{_{d_5}}(U_{_b}^\dagger)_{_{i2}}(W_{_b})_{_{1j}}\cos\alpha\Big]
h^0\overline{b}_{_{i+3}}P_{_R}b_{_{j+3}}
\nonumber\\
&&\hspace{1.8cm}
-\Big[Y_{_{d_4}}(W_{_b}^\dagger)_{_{i2}}(U_{_b})_{_{1j}}\cos\alpha
+Y_{_{d_5}}(W_{_b}^\dagger)_{_{i1}}(U_{_b})_{_{2j}}\sin\alpha\Big]
H^0\overline{b}_{_{i+3}}P_{_L}b_{_{j+3}}
\nonumber\\
&&\hspace{1.8cm}
-\Big[Y_{_{d_4}}(U_{_b}^\dagger)_{_{i1}}(W_{_b})_{_{2j}}\cos\alpha
+Y_{_{d_5}}(U_{_b}^\dagger)_{_{i2}}(W_{_b})_{_{1j}}\sin\alpha\Big]
H^0\overline{b}_{_{i+3}}P_{_R}b_{_{j+3}}
\nonumber\\
&&\hspace{1.8cm}
+{i\over\sqrt{2}}\sum\limits_{i,j=1}^2\Big\{
\Big[Y_{_{d_4}}(W_{_b}^\dagger)_{_{i2}}(U_{_b})_{_{1j}}\sin\beta
-Y_{_{d_5}}(W_{_b}^\dagger)_{_{i1}}(U_{_b})_{_{2j}}\cos\beta\Big]
A^0\overline{b}_{_{i+3}}P_{_L}b_{_{j+3}}
\nonumber\\
&&\hspace{1.8cm}
-\Big[Y_{_{d_4}}(U_{_b}^\dagger)_{_{i1}}(W_{_b})_{_{2j}}\sin\beta
-Y_{_{d_5}}(U_{_b}^\dagger)_{_{i2}}(W_{_b})_{_{1j}}\cos\beta\Big]
A^0\overline{b}_{_{i+3}}P_{_R}b_{_{j+3}}
\nonumber\\
&&\hspace{1.8cm}
-\Big[Y_{_{d_4}}(W_{_b}^\dagger)_{_{i2}}(U_{_b})_{_{1j}}\cos\beta
+Y_{_{d_5}}(W_{_b}^\dagger)_{_{i1}}(U_{_b})_{_{2j}}\sin\beta\Big]
G^0\overline{b}_{_{i+3}}P_{_L}b_{_{j+3}}
\nonumber\\
&&\hspace{1.8cm}
+\Big[Y_{_{d_4}}(U_{_b}^\dagger)_{_{i1}}(W_{_b})_{_{2j}}\cos\beta
+Y_{_{d_5}}(U_{_b}^\dagger)_{_{i2}}(W_{_b})_{_{1j}}\sin\beta\Big]
G^0\overline{b}_{_{i+3}}P_{_R}b_{_{j+3}}
\nonumber\\
&&\hspace{1.8cm}
-{1\over\sqrt{2}}\sum\limits_{i,j=1}^2\Big\{
\Big[\lambda_{_d}(W_{_b}^\dagger)_{_{i2}}(U_{_b})_{_{2j}}\cos\alpha_{_B}
+\lambda_{_Q}(W_{_b}^\dagger)_{_{i1}}(U_{_b})_{_{1j}}\sin\alpha_{_B}\Big]
h_{_B}^0\overline{b}_{_{i+3}}P_{_L}b_{_{j+3}}
\nonumber\\
&&\hspace{1.8cm}
+\Big[\lambda_{_d}(U_{_b}^\dagger)_{_{i2}}(W_{_b})_{_{2j}}\cos\alpha_{_B}
+\lambda_{_Q}(U_{_b}^\dagger)_{_{i2}}(W_{_b})_{_{2j}}\sin\alpha_{_B}\Big]
h_{_B}^0\overline{b}_{_{i+3}}P_{_R}b_{_{j+3}}
\nonumber\\
&&\hspace{1.8cm}
+\Big[\lambda_{_d}(W_{_b}^\dagger)_{_{i2}}(U_{_b})_{_{2j}}\sin\alpha_{_B}
+\lambda_{_Q}(W_{_b}^\dagger)_{_{i1}}(U_{_b})_{_{1j}}\cos\alpha_{_B}\Big]
H_{_B}^0\overline{b}_{_{i+3}}P_{_L}b_{_{j+3}}
\nonumber\\
&&\hspace{1.8cm}
+\Big[\lambda_{_d}(U_{_b}^\dagger)_{_{i2}}(W_{_b})_{_{2j}}\sin\alpha_{_B}
+\lambda_{_Q}(U_{_b}^\dagger)_{_{i2}}(W_{_b})_{_{2j}}\cos\alpha_{_B}\Big]
H_{_B}^0\overline{b}_{_{i+3}}P_{_R}b_{_{j+3}}\Big\}
\nonumber\\
&&\hspace{1.8cm}
-{i\over\sqrt{2}}\sum\limits_{i,j=1}^2\Big\{
\Big[\lambda_{_d}(W_{_b}^\dagger)_{_{i2}}(U_{_b})_{_{2j}}\cos\beta_{_B}
+\lambda_{_Q}(W_{_b}^\dagger)_{_{i1}}(U_{_b})_{_{1j}}\sin\beta_{_B}\Big]
A_{_B}^0\overline{b}_{_{i+3}}P_{_L}b_{_{j+3}}
\nonumber\\
&&\hspace{1.8cm}
-\Big[\lambda_{_d}(U_{_b}^\dagger)_{_{i2}}(W_{_b})_{_{2j}}\cos\beta_{_B}
+\lambda_{_Q}(U_{_b}^\dagger)_{_{i2}}(W_{_b})_{_{2j}}\sin\beta_{_B}\Big]
A_{_B}^0\overline{b}_{_{i+3}}P_{_R}b_{_{j+3}}
\nonumber\\
&&\hspace{1.8cm}
+\Big[\lambda_{_d}(W_{_b}^\dagger)_{_{i2}}(U_{_b})_{_{2j}}\sin\beta_{_B}
+\lambda_{_Q}(W_{_b}^\dagger)_{_{i1}}(U_{_b})_{_{1j}}\cos\beta_{_B}\Big]
G_{_B}^0\overline{b}_{_{i+3}}P_{_L}b_{_{j+3}}
\nonumber\\
&&\hspace{1.8cm}
-\Big[\lambda_{_d}(U_{_b}^\dagger)_{_{i2}}(W_{_b})_{_{2j}}\sin\beta_{_B}
+\lambda_{_Q}(U_{_b}^\dagger)_{_{i2}}(W_{_b})_{_{2j}}\cos\beta_{_B}\Big]
G_{_B}^0\overline{b}_{_{i+3}}P_{_R}b_{_{j+3}}\Big\}
\label{app1-1/3}
\end{eqnarray}

\section{mass squared matrices for exotic squarks\label{app2}}
\indent\indent
For charged $2/3$ exotic scalar quarks, the elements of mass squared matrix are written as
\begin{eqnarray}
&&{\cal M}_{\tilde{t}^\prime}^2(\tilde{Q}_{_4}^{1*}\tilde{Q}_{_4}^1)
=m_{_{\tilde{Q}_4}}^2+{1\over2}Y_{_{u_4}}^2\upsilon_{_u}^2+{1\over2}Y_{_{d_4}}^2\upsilon_{_d}^2
+{1\over2}\lambda_{_Q}^2\upsilon_{_B}^2+\Big({1\over2}-{2\over3}s_{_{\rm W}}^2\Big)m_{_{\rm Z}}^2\cos2\beta
\nonumber\\
&&\hspace{2.7cm}
+{B_{_4}\over2}m_{_{Z_B}}^2\cos2\beta_{_B}\;,
\nonumber\\
&&{\cal M}_{\tilde{t}^\prime}^2(\tilde{U}_{_4}^c\tilde{U}_{_4}^{c*})
=m_{_{\tilde{U}_4}}^2+{1\over2}Y_{_{u_4}}^2\upsilon_{_u}^2
+{1\over2}\lambda_{_u}^2\overline{\upsilon}_{_B}^2-{2\over3}s_{_{\rm W}}^2m_{_{\rm Z}}^2\cos2\beta
-{B_{_4}\over2}m_{_{Z_B}}^2\cos2\beta_{_B}\;,
\nonumber\\
&&{\cal M}_{\tilde{t}^\prime}^2(\tilde{Q}_{_5}^{2c}\tilde{Q}_{_5}^{2c*})
=m_{_{\tilde{Q}_5}}^2+{1\over2}Y_{_{u_5}}^2\upsilon_{_d}^2+{1\over2}Y_{_{d_5}}^2\upsilon_{_u}^2
+{1\over2}\lambda_{_Q}^2\upsilon_{_B}^2+\Big({1\over2}-{1\over3}s_{_{\rm W}}^2\Big)m_{_{\rm Z}}^2\cos2\beta
\nonumber\\
&&\hspace{2.9cm}
-{1+B_{_4}\over2}m_{_{Z_B}}^2\cos2\beta_{_B}\;,
\nonumber\\
&&{\cal M}_{\tilde{t}^\prime}^2(\tilde{U}_{_5}^*\tilde{U}_{_5})
=m_{_{\tilde{U}_5}}^2+{1\over2}Y_{_{u_5}}^2\upsilon_{_d}^2
+{1\over2}\lambda_{_u}^2\overline{\upsilon}_{_B}^2+{2\over3}s_{_{\rm W}}^2m_{_{\rm Z}}^2\cos2\beta
+{1+B_{_4}\over2}m_{_{Z_B}}^2\cos2\beta_{_B}\;,
\nonumber\\
&&{\cal M}_{\tilde{t}^\prime}^2(\tilde{U}_{_4}^c\tilde{Q}_{_4}^1)
=-{1\over\sqrt{2}}\upsilon_{_u}Y_{_{u_4}}A_{_{u_4}}+{1\over\sqrt{2}}Y_{_{u_4}}\mu\upsilon_{_d}\;,
\nonumber\\
&&{\cal M}_{\tilde{t}^\prime}^2(\tilde{Q}_{_5}^{2c}\tilde{Q}_{_4}^1)
=-{1\over\sqrt{2}}\upsilon_{_B}\lambda_{_Q}A_{_{BQ}}+\sqrt{2}\lambda_{_Q}\mu_{_B}\overline{\upsilon}_{_B}\;,
\nonumber\\
&&{\cal M}_{\tilde{t}^\prime}^2(\tilde{U}_{_5}^*\tilde{Q}_{_4}^1)
=-{1\over\sqrt{2}}Y_{_{u_4}}\lambda_u\upsilon_{_u}\overline{\upsilon}_{_B}
+{1\over\sqrt{2}}Y_{_{u_5}}\lambda_Q\upsilon_{_d}\upsilon_{_B}\;,
\nonumber\\
&&{\cal M}_{\tilde{t}^\prime}^2(\tilde{Q}_{_5}^{2c}\tilde{U}_{_4}^{c*})
={1\over2}\lambda_Q Y_{_{u_4}}\upsilon_{_u}\upsilon_{_B}-{1\over2}\lambda_uY_{_{u_5}}
\upsilon_{_d}\overline{\upsilon}_{_B}\;,
\nonumber\\
&&{\cal M}_{\tilde{t}^\prime}^2(\tilde{U}_{_5}^*\tilde{U}_{_4}^{c*})
=-{1\over\sqrt{2}}\lambda_{_u}A_{_{BU}}\overline{\upsilon}_{_B}+{1\over\sqrt{2}}\lambda_{_u}\mu_{_B}
\upsilon_{_B}\;,
\nonumber\\
&&{\cal M}_{\tilde{t}^\prime}^2(\tilde{Q}_{_5}^{2c}\tilde{U}_{_5})=-{1\over\sqrt{2}}
Y_{_{u_5}}A_{_{u_5}}\upsilon_{_d}+{1\over\sqrt{2}}Y_{_{u_5}}\mu\upsilon_{_u}\;.
\label{ESQ-2/3}
\end{eqnarray}

For charged $-1/3$ exotic scalar quarks, the elements of mass squared matrix are given as
\begin{eqnarray}
&&{\cal M}_{\tilde{t}^\prime}^2(\tilde{Q}_{_4}^{2*}\tilde{Q}_{_4}^2)
=m_{_{\tilde{Q}_4}}^2+{1\over2}Y_{_{u_4}}^2\upsilon_{_u}^2+{1\over2}Y_{_{d_4}}^2\upsilon_{_d}^2
+{1\over2}\lambda_{_Q}^2\upsilon_{_B}^2-\Big({1\over2}-{2\over3}s_{_{\rm W}}^2\Big)m_{_{\rm Z}}^2\cos2\beta
\nonumber\\
&&\hspace{2.7cm}
+{B_{_4}\over2}m_{_{Z_B}}^2\cos2\beta_{_B}\;,
\nonumber\\
&&{\cal M}_{\tilde{t}^\prime}^2(\tilde{D}_{_4}^c\tilde{D}_{_4}^{c*})
=m_{_{\tilde{D}_4}}^2+{1\over2}Y_{_{d_4}}^2\upsilon_{_d}^2
+{1\over2}\lambda_{_d}^2\overline{\upsilon}_{_B}^2-{1\over3}s_{_{\rm W}}^2m_{_{\rm Z}}^2\cos2\beta
-{B_{_4}\over2}m_{_{Z_B}}^2\cos2\beta_{_B}\;,
\nonumber\\
&&{\cal M}_{\tilde{t}^\prime}^2(\tilde{Q}_{_5}^{1c}\tilde{Q}_{_5}^{1c*})
=m_{_{\tilde{Q}_5}}^2+{1\over2}Y_{_{u_5}}^2\upsilon_{_d}^2+{1\over2}Y_{_{d_5}}^2\upsilon_{_u}^2
+{1\over2}\lambda_{_Q}^2\upsilon_{_B}^2-\Big({1\over2}+{1\over3}s_{_{\rm W}}^2\Big)m_{_{\rm Z}}^2\cos2\beta
\nonumber\\
&&\hspace{2.9cm}
-{1+B_{_4}\over2}m_{_{Z_B}}^2\cos2\beta_{_B}\;,
\nonumber\\
&&{\cal M}_{\tilde{t}^\prime}^2(\tilde{D}_{_5}^*\tilde{D}_{_5})
=m_{_{\tilde{D}_5}}^2+{1\over2}Y_{_{d_5}}^2\upsilon_{_u}^2
+{1\over2}\lambda_{_d}^2\overline{\upsilon}_{_B}^2+{1\over3}s_{_{\rm W}}^2m_{_{\rm Z}}^2\cos2\beta
+{1+B_{_4}\over2}m_{_{Z_B}}^2\cos2\beta_{_B}\;,
\nonumber\\
&&{\cal M}_{\tilde{t}^\prime}^2(\tilde{D}_{_4}^c\tilde{Q}_{_4}^2)
=-{1\over\sqrt{2}}Y_{_{d_4}}\upsilon_{_d}A_{_{d_4}}+{1\over\sqrt{2}}Y_{_{d_4}}\mu\upsilon_{_d}\;,
\nonumber\\
&&{\cal M}_{\tilde{t}^\prime}^2(\tilde{Q}_{_5}^{1c}\tilde{Q}_{_4}^2)
=-{1\over\sqrt{2}}\lambda_{_Q}\upsilon_{_B}A_{_{BQ}}+\sqrt{2}\lambda_{_Q}\mu_{_B}\overline{\upsilon}_{_B}\;,
\nonumber\\
&&{\cal M}_{\tilde{t}^\prime}^2(\tilde{D}_{_5}^*\tilde{Q}_{_4}^2)
=-{1\over\sqrt{2}}Y_{_{d_4}}\lambda_d\upsilon_{_d}\overline{\upsilon}_{_B}
+{1\over\sqrt{2}}Y_{_{d_5}}\lambda_Q\upsilon_{_u}\upsilon_{_B}\;,
\nonumber\\
&&{\cal M}_{\tilde{t}^\prime}^2(\tilde{Q}_{_5}^{1c}\tilde{D}_{_4}^{c*})
={1\over2}\lambda_Q Y_{_{d_4}}\upsilon_{_d}\upsilon_{_B}+{1\over2}\lambda_dY_{_{d_5}}
\upsilon_{_u}\overline{\upsilon}_{_B}\;,
\nonumber\\
&&{\cal M}_{\tilde{t}^\prime}^2(\tilde{D}_{_5}^*\tilde{D}_{_4}^{c*})
=-{1\over\sqrt{2}}\lambda_{_d}A_{_{BD}}\overline{\upsilon}_{_B}+{1\over\sqrt{2}}\lambda_{_d}\mu_{_B}
\upsilon_{_B}\;,
\nonumber\\
&&{\cal M}_{\tilde{t}^\prime}^2(\tilde{Q}_{_5}^{1c}\tilde{D}_{_5})=-{1\over\sqrt{2}}
Y_{_{d_5}}A_{_{d_5}}\upsilon_{_u}+{1\over\sqrt{2}}Y_{_{d_5}}\mu\upsilon_{_d}\;.
\label{ESQ-1/3}
\end{eqnarray}

\section{The couplings between neutral Higgs and exotic squarks\label{app3}}
\indent\indent
In the mass basis, the couplings between the neutral Higgs and exotic squarks are
\begin{eqnarray}
&&{\cal L}_{_{H\tilde{\cal U}_i^*\tilde{\cal U}_j}}=\sum\limits_{i,j}^4\Big\{\Big[\xi_{_{uij}}^S\cos\alpha
-\xi_{_{dij}}^S\sin\alpha\Big]h^0\tilde{\cal U}_i^*\tilde{\cal U}_j+\Big[\eta_{_{uij}}^S\cos\alpha
-\eta_{_{dij}}^S\sin\alpha\Big]h^0\tilde{\cal D}_i^*\tilde{\cal D}_j
\nonumber\\
&&\hspace{1.7cm}
+\Big[\xi_{_{uij}}^S\sin\alpha
+\xi_{_{dij}}^S\cos\alpha\Big]H^0\tilde{\cal U}_i^*\tilde{\cal U}_j+\Big[\eta_{_{uij}}^S\sin\alpha
+\eta_{_{dij}}^S\cos\alpha\Big]H^0\tilde{\cal D}_i^*\tilde{\cal D}_j
\nonumber\\
&&\hspace{1.7cm}
+i\Big[\xi_{_{uij}}^P\cos\beta
-\xi_{_{dij}}^P\sin\beta\Big]A^0\tilde{\cal U}_i^*\tilde{\cal U}_j+i\Big[\eta_{_{uij}}^P\cos\beta
-\eta_{_{dij}}^P\sin\beta\Big]A^0\tilde{\cal D}_i^*\tilde{\cal D}_j
\nonumber\\
&&\hspace{1.7cm}
+i\Big[\xi_{_{uij}}^P\sin\beta
+\xi_{_{dij}}^P\cos\beta\Big]G^0\tilde{\cal U}_i^*\tilde{\cal U}_j+i\Big[\eta_{_{uij}}^P\sin\beta
+\eta_{_{dij}}^P\cos\beta\Big]G^0\tilde{\cal D}_i^*\tilde{\cal D}_j
\nonumber\\
&&\hspace{1.7cm}
+\Big[\varsigma_{_{uij}}^S\cos\alpha_{_B}
-\varsigma_{_{dij}}^S\sin\alpha_{_B}\Big]h_{_B}^0\tilde{\cal U}_i^*\tilde{\cal U}_j+\Big[\zeta_{_{uij}}^S\cos\alpha_{_B}
-\zeta_{_{dij}}^S\sin\alpha_{_B}\Big]h_{_B}^0\tilde{\cal D}_i^*\tilde{\cal D}_j
\nonumber\\
&&\hspace{1.7cm}
+\Big[\varsigma_{_{uij}}^S\sin\alpha_{_B}
+\varsigma_{_{dij}}^S\cos\alpha_{_B}\Big]H_{_B}^0\tilde{\cal U}_i^*\tilde{\cal U}_j+\Big[\zeta_{_{uij}}^S\sin\alpha_{_B}
+\zeta_{_{dij}}^S\cos\alpha_{_B}\Big]H_{_B}^0\tilde{\cal D}_i^*\tilde{\cal D}_j
\nonumber\\
&&\hspace{1.7cm}
+i\Big[\varsigma_{_{uij}}^P\cos\beta_{_B}
-\varsigma_{_{dij}}^P\sin\beta_{_B}\Big]A_{_B}^0\tilde{\cal U}_i^*\tilde{\cal U}_j+i\Big[\zeta_{_{uij}}^P\cos\beta_{_B}
-\zeta_{_{dij}}^P\sin\beta_{_B}\Big]A_{_B}^0\tilde{\cal D}_i^*\tilde{\cal D}_j
\nonumber\\
&&\hspace{1.7cm}
+i\Big[\varsigma_{_{uij}}^P\sin\beta_{_B}
+\varsigma_{_{dij}}^P\cos\beta_{_B}\Big]G_{_B}^0\tilde{\cal U}_i^*\tilde{\cal U}_j+i\Big[\zeta_{_{uij}}^P\sin\beta_{_B}
+\zeta_{_{dij}}^P\cos\beta_{_B}\Big]G_{_B}^0\tilde{\cal D}_i^*\tilde{\cal D}_j\Big\}\;,
\label{H-EQ}
\end{eqnarray}
with
\begin{eqnarray}
&&\xi_{_{uij}}^S={1\over\sqrt{2}}Y_{_{u_5}}\mu\Big(U_{_{i3}}^\dagger U_{_{4j}}+U_{_{i4}}^\dagger U_{_{3j}}\Big)
+{1\over2}\lambda_{_Q}Y_{_{u_4}}\upsilon_{_B}\Big(U_{_{i3}}^\dagger U_{_{2j}}+U_{_{i2}}^\dagger U_{_{3j}}\Big)
\nonumber\\
&&\hspace{1.2cm}
-{1\over2}\lambda_{_u}Y_{_{u_4}}\overline{\upsilon}_{_B}\Big(U_{_{i1}}^\dagger U_{_{4j}}+U_{_{i4}}^\dagger U_{_{1j}}\Big)
+{e^2\over4s_{_{\rm W}}^2}\upsilon_{_u}\Big(U_{_{i3}}^\dagger U_{_{3j}}-U_{_{i1}}^\dagger U_{_{1j}}\Big)
\nonumber\\
&&\hspace{1.2cm}
+{e^2\over12c_{_{\rm W}}^2}\upsilon_{_u}\Big(U_{_{i1}}^\dagger U_{_{1j}}-U_{_{i3}}^\dagger U_{_{3j}}
-4U_{_{i2}}^\dagger U_{_{2j}}+4U_{_{i4}}^\dagger U_{_{4j}}\Big)
\nonumber\\
&&\hspace{1.2cm}
-{1\over\sqrt{2}}A_{_{u_4}}Y_{_{u_4}}\Big(U_{_{i2}}^\dagger U_{_{1j}}+U_{_{i1}}^\dagger U_{_{2j}}\Big)
\;,\nonumber\\
%%%%%%%%%%%%%%%%%%%%%%%%%%%%%%%%%%%%%%%%%%%%%%%%%%%%%%%%%%%
&&\xi_{_{dij}}^S={1\over\sqrt{2}}Y_{_{u_4}}\mu\Big(U_{_{i2}}^\dagger U_{_{1j}}+U_{_{i1}}^\dagger U_{_{2j}}\Big)
+{1\over2}\lambda_{_Q}Y_{_{u_5}}\upsilon_{_B}\Big(U_{_{i5}}^\dagger U_{_{1j}}+U_{_{i1}}^\dagger U_{_{5j}}\Big)
\nonumber\\
&&\hspace{1.2cm}
-{1\over2}\lambda_{_u}Y_{_{u_5}}\overline{\upsilon}_{_B}\Big(U_{_{i2}}^\dagger U_{_{3j}}+U_{_{i3}}^\dagger U_{_{2j}}\Big)
-{e^2\over4s_{_{\rm W}}^2}\upsilon_{_d}\Big(U_{_{i3}}^\dagger U_{_{3j}}+U_{_{i1}}^\dagger U_{_{1j}}\Big)
\nonumber\\
&&\hspace{1.2cm}
-{e^2\over12c_{_{\rm W}}^2}\upsilon_{_d}\Big(U_{_{i1}}^\dagger U_{_{1j}}-U_{_{i3}}^\dagger U_{_{3j}}
-4U_{_{i2}}^\dagger U_{_{2j}}+4U_{_{i4}}^\dagger U_{_{4j}}\Big)
\nonumber\\
&&\hspace{1.2cm}
-{1\over\sqrt{2}}A_{_{u_5}}Y_{_{u_5}}\Big(U_{_{i3}}^\dagger U_{_{4j}}+U_{_{i4}}^\dagger U_{_{3j}}\Big)
\;,\nonumber\\
%%%%%%%%%%%%%%%%%%%%%%%%%%%%%%%%%%%%%%%%%%%%%%%%%%%%%%%%%%%
&&\eta_{_{uij}}^S={1\over\sqrt{2}}Y_{_{d_4}}\mu\Big(D_{_{i2}}^\dagger D_{_{1j}}+D_{_{i1}}^\dagger D_{_{2j}}\Big)
+{1\over2}\lambda_{_Q}Y_{_{d_5}}\upsilon_{_B}\Big(D_{_{i4}}^\dagger D_{_{1j}}+D_{_{i1}}^\dagger D_{_{4j}}\Big)
\nonumber\\
&&\hspace{1.2cm}
-{1\over2}\lambda_{_d}Y_{_{d_5}}\overline{\upsilon}_{_B}\Big(D_{_{i2}}^\dagger D_{_{3j}}+D_{_{i3}}^\dagger D_{_{2j}}\Big)
+{e^2\over4s_{_{\rm W}}^2}\upsilon_{_u}\Big(D_{_{i1}}^\dagger D_{_{1j}}-D_{_{i3}}^\dagger D_{_{3j}}\Big)
\nonumber\\
&&\hspace{1.2cm}
+{e^2\over12c_{_{\rm W}}^2}\upsilon_{_u}\Big(D_{_{i1}}^\dagger D_{_{1j}}-D_{_{i3}}^\dagger D_{_{3j}}
+2D_{_{i2}}^\dagger D_{_{2j}}-2D_{_{i4}}^\dagger D_{_{4j}}\Big)
\nonumber\\
&&\hspace{1.2cm}
-{1\over\sqrt{2}}A_{_{d_5}}Y_{_{d_5}}\Big(D_{_{i3}}^\dagger D_{_{4j}}+D_{_{i4}}^\dagger D_{_{3j}}\Big)
\;,\nonumber\\
%%%%%%%%%%%%%%%%%%%%%%%%%%%%%%%%%%%%%%%%%%%%%%%%%%%%%%%%%%%
&&\eta_{_{dij}}^S={1\over\sqrt{2}}Y_{_{d_5}}\mu\Big(D_{_{i3}}^\dagger D_{_{4j}}+D_{_{i4}}^\dagger D_{_{3j}}\Big)
+{1\over2}\lambda_{_Q}Y_{_{d_4}}\upsilon_{_B}\Big(D_{_{i3}}^\dagger D_{_{2j}}+D_{_{i2}}^\dagger D_{_{3j}}\Big)
\nonumber\\
&&\hspace{1.2cm}
-{1\over2}\lambda_{_d}Y_{_{d_4}}\overline{\upsilon}_{_B}\Big(D_{_{i1}}^\dagger D_{_{4j}}+D_{_{i4}}^\dagger D_{_{1j}}\Big)
-{e^2\over4s_{_{\rm W}}^2}\upsilon_{_d}\Big(D_{_{i1}}^\dagger D_{_{1j}}-D_{_{i3}}^\dagger D_{_{3j}}\Big)
\nonumber\\
&&\hspace{1.2cm}
-{e^2\over12c_{_{\rm W}}^2}\upsilon_{_u}\Big(D_{_{i1}}^\dagger D_{_{1j}}-D_{_{i3}}^\dagger D_{_{3j}}
+2D_{_{i2}}^\dagger D_{_{2j}}-2D_{_{i4}}^\dagger D_{_{4j}}\Big)
\nonumber\\
&&\hspace{1.2cm}
-{1\over\sqrt{2}}A_{_{d_4}}Y_{_{d_4}}\Big(D_{_{i2}}^\dagger D_{_{1j}}+D_{_{i1}}^\dagger D_{_{2j}}\Big)
\;,\nonumber\\
%%%%%%%%%%%%%%%%%%%%%%%%%%%%%%%%%%%%%%%%%%%%%%%%%%%%%%%%%%%
&&\xi_{_{uij}}^P={1\over\sqrt{2}}Y_{_{u_5}}\mu\Big(U_{_{i3}}^\dagger U_{_{4j}}-U_{_{i4}}^\dagger U_{_{3j}}\Big)
-{1\over2}\lambda_{_Q}Y_{_{u_4}}\upsilon_{_B}\Big(U_{_{i3}}^\dagger U_{_{2j}}-U_{_{i2}}^\dagger U_{_{3j}}\Big)
\nonumber\\
&&\hspace{1.2cm}
+{1\over2}\lambda_{_u}Y_{_{u_4}}\overline{\upsilon}_{_B}\Big(U_{_{i1}}^\dagger U_{_{4j}}-U_{_{i4}}^\dagger U_{_{1j}}\Big)
-{1\over\sqrt{2}}A_{_{u_4}}Y_{_{u_4}}\Big(U_{_{i2}}^\dagger U_{_{1j}}-U_{_{i1}}^\dagger U_{_{2j}}\Big)
\;,\nonumber\\
%%%%%%%%%%%%%%%%%%%%%%%%%%%%%%%%%%%%%%%%%%%%%%%%%%%%%%%%%%%
&&\xi_{_{dij}}^P={1\over\sqrt{2}}Y_{_{u_4}}\mu\Big(U_{_{i2}}^\dagger U_{_{1j}}-U_{_{i1}}^\dagger U_{_{2j}}\Big)
-{1\over2}\lambda_{_Q}Y_{_{u_5}}\upsilon_{_B}\Big(U_{_{i5}}^\dagger U_{_{1j}}-U_{_{i1}}^\dagger U_{_{5j}}\Big)
\nonumber\\
&&\hspace{1.2cm}
+{1\over2}\lambda_{_u}Y_{_{u_5}}\overline{\upsilon}_{_B}\Big(U_{_{i2}}^\dagger U_{_{3j}}-U_{_{i3}}^\dagger U_{_{2j}}\Big)
-{1\over\sqrt{2}}A_{_{u_5}}Y_{_{u_5}}\Big(U_{_{i3}}^\dagger U_{_{4j}}-U_{_{i4}}^\dagger U_{_{3j}}\Big)
\;,\nonumber\\
%%%%%%%%%%%%%%%%%%%%%%%%%%%%%%%%%%%%%%%%%%%%%%%%%%%%%%%%%%%
&&\eta_{_{uij}}^P={1\over\sqrt{2}}Y_{_{d_4}}\mu\Big(D_{_{i2}}^\dagger D_{_{1j}}-D_{_{i1}}^\dagger D_{_{2j}}\Big)
-{1\over2}\lambda_{_Q}Y_{_{d_5}}\upsilon_{_B}\Big(D_{_{i4}}^\dagger D_{_{1j}}-D_{_{i1}}^\dagger D_{_{4j}}\Big)
\nonumber\\
&&\hspace{1.2cm}
+{1\over2}\lambda_{_d}Y_{_{d_5}}\overline{\upsilon}_{_B}\Big(D_{_{i2}}^\dagger D_{_{3j}}-D_{_{i3}}^\dagger D_{_{2j}}\Big)
-{1\over\sqrt{2}}A_{_{d_5}}Y_{_{d_5}}\Big(D_{_{i3}}^\dagger D_{_{4j}}-D_{_{i4}}^\dagger D_{_{3j}}\Big)
\;,\nonumber\\
%%%%%%%%%%%%%%%%%%%%%%%%%%%%%%%%%%%%%%%%%%%%%%%%%%%%%%%%%%%
&&\eta_{_{dij}}^P={1\over\sqrt{2}}Y_{_{d_5}}\mu\Big(D_{_{i3}}^\dagger D_{_{4j}}-D_{_{i4}}^\dagger D_{_{3j}}\Big)
-{1\over2}\lambda_{_Q}Y_{_{d_4}}\upsilon_{_B}\Big(D_{_{i3}}^\dagger D_{_{2j}}-D_{_{i2}}^\dagger D_{_{3j}}\Big)
\nonumber\\
&&\hspace{1.2cm}
+{1\over2}\lambda_{_d}Y_{_{d_4}}\overline{\upsilon}_{_B}\Big(D_{_{i1}}^\dagger D_{_{4j}}-D_{_{i4}}^\dagger D_{_{1j}}\Big)
-{1\over\sqrt{2}}A_{_{d_4}}Y_{_{d_4}}\Big(D_{_{i2}}^\dagger D_{_{1j}}-D_{_{i1}}^\dagger D_{_{2j}}\Big)
\;,\nonumber\\
%%%%%%%%%%%%%%%%%%%%%%%%%%%%%%%%%%%%%%%%%%%%%%%%%%%%%%%%%%%
&&\varsigma_{_{uij}}^S={1\over\sqrt{2}}\lambda_{_u}\mu_{_B}\Big(U_{_{i2}}^\dagger U_{_{4j}}+U_{_{i4}}^\dagger U_{_{2j}}\Big)
+{1\over2}\lambda_{_Q}Y_{_{u_4}}\upsilon_{_u}\Big(U_{_{i3}}^\dagger U_{_{2j}}+U_{_{i2}}^\dagger U_{_{3j}}\Big)
\nonumber\\
&&\hspace{1.2cm}
-{1\over2}\lambda_{_Q}Y_{_{u_5}}\upsilon_{_d}\Big(U_{_{i4}}^\dagger U_{_{1j}}+U_{_{i1}}^\dagger U_{_{4j}}\Big)
+g_{_B}^2\upsilon_{_B}\Big(B_{_4}U_{_{i1}}^\dagger U_{_{1j}}-(1+B_{_4})U_{_{i3}}^\dagger U_{_{3j}}
\nonumber\\
&&\hspace{1.2cm}
-B_{_4}U_{_{i2}}^\dagger U_{_{2j}}+(1+B_{_4})U_{_{i4}}^\dagger U_{_{4j}}\Big)
-{1\over\sqrt{2}}A_{_{BQ}}\lambda_{_Q}\Big(U_{_{i3}}^\dagger U_{_{1j}}+U_{_{i1}}^\dagger U_{_{3j}}\Big)
\;,\nonumber\\
%%%%%%%%%%%%%%%%%%%%%%%%%%%%%%%%%%%%%%%%%%%%%%%%%%%%%%%%%%%
&&\varsigma_{_{dij}}^S={1\over\sqrt{2}}\lambda_{_Q}\mu_{_B}\Big(U_{_{i3}}^\dagger U_{_{1j}}+U_{_{i1}}^\dagger U_{_{3j}}\Big)
-{1\over2}\lambda_{_u}Y_{_{u_4}}\upsilon_{_u}\Big(U_{_{i1}}^\dagger U_{_{4j}}+U_{_{i4}}^\dagger U_{_{1j}}\Big)
\nonumber\\
&&\hspace{1.2cm}
-{1\over2}\lambda_{_u}Y_{_{u_5}}\upsilon_{_d}\Big(U_{_{i2}}^\dagger U_{_{3j}}+U_{_{i3}}^\dagger U_{_{2j}}\Big)
-g_{_B}^2\overline{\upsilon}_{_B}\Big(B_{_4}U_{_{i1}}^\dagger U_{_{1j}}-(1+B_{_4})U_{_{i3}}^\dagger U_{_{3j}}
\nonumber\\
&&\hspace{1.2cm}
-B_{_4}U_{_{i2}}^\dagger U_{_{2j}}+(1+B_{_4})U_{_{i4}}^\dagger U_{_{4j}}\Big)
+{1\over\sqrt{2}}A_{_{BU}}\lambda_{_u}\Big(U_{_{i2}}^\dagger U_{_{4j}}+U_{_{i4}}^\dagger U_{_{2j}}\Big)
\;,\nonumber\\
%%%%%%%%%%%%%%%%%%%%%%%%%%%%%%%%%%%%%%%%%%%%%%%%%%%%%%%%%%%
&&\zeta_{_{uij}}^S={1\over\sqrt{2}}\lambda_{_d}\mu_{_B}\Big(D_{_{i2}}^\dagger D_{_{4j}}+D_{_{i4}}^\dagger D_{_{2j}}\Big)
+{1\over2}\lambda_{_Q}Y_{_{d_4}}\upsilon_{_d}\Big(D_{_{i3}}^\dagger D_{_{2j}}+D_{_{i2}}^\dagger D_{_{3j}}\Big)
\nonumber\\
&&\hspace{1.2cm}
-{1\over2}\lambda_{_Q}Y_{_{d_5}}\upsilon_{_u}\Big(D_{_{i4}}^\dagger D_{_{1j}}+D_{_{i1}}^\dagger D_{_{4j}}\Big)
+g_{_B}^2\upsilon_{_B}\Big(B_{_4}D_{_{i1}}^\dagger D_{_{1j}}-(1+B_{_4})D_{_{i3}}^\dagger D_{_{3j}}
\nonumber\\
&&\hspace{1.2cm}
-B_{_4}D_{_{i2}}^\dagger D_{_{2j}}+(1+B_{_4})D_{_{i4}}^\dagger D_{_{4j}}\Big)
-{1\over\sqrt{2}}A_{_{BQ}}\lambda_{_Q}\Big(D_{_{i3}}^\dagger D_{_{1j}}+D_{_{i1}}^\dagger D_{_{3j}}\Big)
\;,\nonumber\\
%%%%%%%%%%%%%%%%%%%%%%%%%%%%%%%%%%%%%%%%%%%%%%%%%%%%%%%%%%%
&&\zeta_{_{dij}}^S=-{1\over\sqrt{2}}\lambda_{_Q}\mu_{_B}\Big(D_{_{i3}}^\dagger D_{_{1j}}+D_{_{i1}}^\dagger D_{_{3j}}\Big)
-{1\over2}\lambda_{_d}Y_{_{d_4}}\upsilon_{_d}\Big(D_{_{i1}}^\dagger D_{_{4j}}+D_{_{i4}}^\dagger D_{_{1j}}\Big)
\nonumber\\
&&\hspace{1.2cm}
-{1\over2}\lambda_{_d}Y_{_{d_5}}\upsilon_{_u}\Big(D_{_{i2}}^\dagger D_{_{3j}}+D_{_{i3}}^\dagger D_{_{2j}}\Big)
-g_{_B}^2\overline{\upsilon}_{_B}\Big(B_{_4}D_{_{i1}}^\dagger D_{_{1j}}-(1+B_{_4})D_{_{i3}}^\dagger D_{_{3j}}
\nonumber\\
&&\hspace{1.2cm}
-B_{_4}D_{_{i2}}^\dagger D_{_{2j}}+(1+B_{_4})D_{_{i4}}^\dagger D_{_{4j}}\Big)
+{1\over\sqrt{2}}A_{_{BD}}\lambda_{_d}\Big(D_{_{i2}}^\dagger D_{_{4j}}+D_{_{i4}}^\dagger D_{_{2j}}\Big)
\;,\nonumber\\
%%%%%%%%%%%%%%%%%%%%%%%%%%%%%%%%%%%%%%%%%%%%%%%%%%%%%%%%%%%
&&\varsigma_{_{uij}}^P={1\over\sqrt{2}}\lambda_{_u}\mu_{_B}\Big(U_{_{i2}}^\dagger U_{_{4j}}-U_{_{i4}}^\dagger U_{_{2j}}\Big)
+{1\over2}\lambda_{_Q}Y_{_{u_4}}\upsilon_{_u}\Big(U_{_{i3}}^\dagger U_{_{2j}}-U_{_{i2}}^\dagger U_{_{3j}}\Big)
\nonumber\\
&&\hspace{1.2cm}
-{1\over2}\lambda_{_Q}Y_{_{u_5}}\upsilon_{_d}\Big(U_{_{i4}}^\dagger U_{_{1j}}-U_{_{i1}}^\dagger U_{_{4j}}\Big)
-{1\over\sqrt{2}}A_{_{BQ}}\lambda_{_Q}\Big(U_{_{i3}}^\dagger U_{_{1j}}-U_{_{i1}}^\dagger U_{_{3j}}\Big)
\;,\nonumber\\
%%%%%%%%%%%%%%%%%%%%%%%%%%%%%%%%%%%%%%%%%%%%%%%%%%%%%%%%%%%
&&\varsigma_{_{dij}}^P={1\over\sqrt{2}}\lambda_{_Q}\mu_{_B}\Big(U_{_{i3}}^\dagger U_{_{1j}}-U_{_{i1}}^\dagger U_{_{3j}}\Big)
-{1\over2}\lambda_{_u}Y_{_{u_4}}\upsilon_{_u}\Big(U_{_{i1}}^\dagger U_{_{4j}}-U_{_{i4}}^\dagger U_{_{1j}}\Big)
\nonumber\\
&&\hspace{1.2cm}
-{1\over2}\lambda_{_u}Y_{_{u_5}}\upsilon_{_d}\Big(U_{_{i2}}^\dagger U_{_{3j}}-U_{_{i3}}^\dagger U_{_{2j}}\Big)
+{1\over\sqrt{2}}A_{_{BU}}\lambda_{_u}\Big(U_{_{i2}}^\dagger U_{_{4j}}-U_{_{i4}}^\dagger U_{_{2j}}\Big)
\;,\nonumber\\
%%%%%%%%%%%%%%%%%%%%%%%%%%%%%%%%%%%%%%%%%%%%%%%%%%%%%%%%%%%
&&\zeta_{_{uij}}^P={1\over\sqrt{2}}\lambda_{_d}\mu_{_B}\Big(D_{_{i2}}^\dagger D_{_{4j}}-D_{_{i4}}^\dagger D_{_{2j}}\Big)
+{1\over2}\lambda_{_Q}Y_{_{d_4}}\upsilon_{_d}\Big(D_{_{i3}}^\dagger D_{_{2j}}-D_{_{i2}}^\dagger D_{_{3j}}\Big)
\nonumber\\
&&\hspace{1.2cm}
-{1\over2}\lambda_{_Q}Y_{_{d_5}}\upsilon_{_u}\Big(D_{_{i4}}^\dagger D_{_{1j}}-D_{_{i1}}^\dagger D_{_{4j}}\Big)
-{1\over\sqrt{2}}A_{_{BQ}}\lambda_{_Q}\Big(D_{_{i3}}^\dagger D_{_{1j}}-D_{_{i1}}^\dagger D_{_{3j}}\Big)
\;,\nonumber\\
%%%%%%%%%%%%%%%%%%%%%%%%%%%%%%%%%%%%%%%%%%%%%%%%%%%%%%%%%%%
&&\zeta_{_{dij}}^P=-{1\over\sqrt{2}}\lambda_{_Q}\mu_{_B}\Big(D_{_{i3}}^\dagger D_{_{1j}}-D_{_{i1}}^\dagger D_{_{3j}}\Big)
-{1\over2}\lambda_{_d}Y_{_{d_4}}\upsilon_{_d}\Big(D_{_{i1}}^\dagger D_{_{4j}}-D_{_{i4}}^\dagger D_{_{1j}}\Big)
\nonumber\\
&&\hspace{1.2cm}
-{1\over2}\lambda_{_d}Y_{_{d_5}}\upsilon_{_u}\Big(D_{_{i2}}^\dagger D_{_{3j}}-D_{_{i3}}^\dagger D_{_{2j}}\Big)
+{1\over\sqrt{2}}A_{_{BD}}\lambda_{_d}\Big(D_{_{i2}}^\dagger D_{_{4j}}-D_{_{i4}}^\dagger D_{_{2j}}\Big)\;.
\label{xi-eta}
\end{eqnarray}

\section{The radiative corrections to the mass squared matrix from exotic lepton fields\label{app4}}
\indent\indent
\begin{eqnarray}
%%%%%%%%%%%%%%%%%%%%%%%%%%%%%%%%%%%%%%%%%%%%%%%%%%%%%%%%%%
&&\Delta_{11}^{L}={G_{_F}m_{_{\nu_4}}^4\over\sqrt{2}\pi^2\sin^2\beta}\cdot
{\mu^2(A_{_{\nu_4}}-\mu\cot\beta)^2\over(m_{_{\tilde{\nu}_4^1}}^2-m_{_{\tilde{\nu}_4^2}}^2)^2}
g(m_{_{\tilde{\nu}_4^1}},m_{_{\tilde{\nu}_4^2}})
\nonumber\\
&&\hspace{1.2cm}
+{G_{_F}m_{_{e_4}}^4\over\sqrt{2}\pi^2\cos^2\beta}\Big\{\ln{m_{_{\tilde{e}_4^1}}m_{_{\tilde{e}_4^2}}
\over m_{_{e_4}}^2}+{A_{_{e_4}}(A_{_{e_4}}-\mu\tan\beta)\over m_{_{\tilde{e}_4^1}}^2-m_{_{\tilde{e}_4^2}}^2}
\ln{m_{_{\tilde{e}_4^1}}^2\over m_{_{\tilde{e}_4^2}}^2}
\nonumber\\
&&\hspace{1.2cm}
+{A_{_{e_4}}^2(A_{_{e_4}}-\mu\tan\beta)^2\over(m_{_{\tilde{e}_4^1}}^2-m_{_{\tilde{e}_4^2}}^2)^2}
g(m_{_{\tilde{e}_4^1}}, m_{_{\tilde{e}_4^2}})\Big\}
\nonumber\\
&&\hspace{1.2cm}
+{G_{_F}m_{_{\nu_5}}^4\over\sqrt{2}\pi^2\cos^2\beta}\Big\{\ln{m_{_{\tilde{\nu}_5^1}}m_{_{\tilde{\nu}_5^2}}
\over m_{_{\nu_5}}^2}+{A_{_{\nu_5}}(A_{_{\nu_5}}-\mu\tan\beta)\over m_{_{\tilde{\nu}_5^1}}^2-m_{_{\tilde{\nu}_5^2}}^2}
\ln{m_{_{\tilde{\nu}_5^1}}^2\over m_{_{\tilde{\nu}_5^2}}^2}
\nonumber\\
&&\hspace{1.2cm}
+{A_{_{\nu_5}}^2(A_{_{\nu_5}}-\mu\tan\beta)^2\over(m_{_{\tilde{\nu}_5^1}}^2-m_{_{\tilde{\nu}_5^2}}^2)^2}
g(m_{_{\tilde{\nu}_5^1}}, m_{_{\tilde{\nu}_5^2}})\Big\}
\nonumber\\
&&\hspace{1.2cm}
+{G_{_F}m_{_{e_5}}^4\over\sqrt{2}\pi^2\sin^2\beta}\cdot
{\mu^2(A_{_{e_5}}-\mu\cot\beta)^2\over(m_{_{\tilde{e}_5^1}}^2-m_{_{\tilde{e}_5^2}}^2)^2}
g(m_{_{\tilde{e}_5^1}},m_{_{\tilde{e}_5^2}})
\;,\nonumber\\
%%%%%%%%%%%%%%%%%%%%%%%%%%%%%%%%%%%%%%%%%%%%%%%%%%%%%%%%%%
&&\Delta_{12}^{L}={G_{_F}m_{_{\nu_4}}^4\over2\sqrt{2}\pi^2\sin^2\beta}\cdot
{\mu(-A_{_{\nu_4}}+\mu\cot\beta)\over m_{_{\tilde{\nu}_4^1}}^2-m_{_{\tilde{\nu}_4^2}}^2}
\Big\{\ln{m_{_{\tilde{\nu}_4^1}}\over m_{_{\tilde{\nu}_4^2}}}+{A_{_{\nu_4}}(A_{_{\nu_4}}-\mu\cot\beta)
\over m_{_{\tilde{\nu}_4^1}}^2-m_{_{\tilde{\nu}_4^2}}^2}g(m_{_{\tilde{\nu}_4^1}},m_{_{\tilde{\nu}_4^2}})\Big\}
\nonumber\\
&&\hspace{1.2cm}
+{G_{_F}m_{_{e_4}}^4\over2\sqrt{2}\pi^2\cos^2\beta}\cdot
{\mu(-A_{_{e_4}}+\mu\tan\beta)\over m_{_{\tilde{e}_4^1}}^2-m_{_{\tilde{e}_4^2}}^2}
\Big\{\ln{m_{_{\tilde{e}_4^1}}\over m_{_{\tilde{e}_4^2}}}+{A_{_{e_4}}(A_{_{e_4}}-\mu\tan\beta)
\over m_{_{\tilde{e}_4^1}}^2-m_{_{\tilde{e}_4^2}}^2}g(m_{_{\tilde{e}_4^1}},m_{_{\tilde{e}_4^2}})\Big\}
\nonumber\\
&&\hspace{1.2cm}
+{G_{_F}m_{_{\nu_5}}^4\over2\sqrt{2}\pi^2\cos^2\beta}\cdot
{\mu(-A_{_{\nu_5}}+\mu\tan\beta)\over m_{_{\tilde{\nu}_5^1}}^2-m_{_{\tilde{\nu}_5^2}}^2}
\Big\{\ln{m_{_{\tilde{\nu}_5^1}}\over m_{_{\tilde{\nu}_5^2}}}+{A_{_{\nu_5}}(A_{_{\nu_5}}-\mu\tan\beta)
\over m_{_{\tilde{\nu}_5^1}}^2-m_{_{\tilde{\nu}_5^2}}^2}g(m_{_{\tilde{\nu}_5^1}},m_{_{\tilde{\nu}_5^2}})\Big\}
\nonumber\\
&&\hspace{1.2cm}
+{G_{_F}m_{_{e_5}}^4\over2\sqrt{2}\pi^2\sin^2\beta}\cdot
{\mu(-A_{_{e_5}}+\mu\cot\beta)\over m_{_{\tilde{e}_5^1}}^2-m_{_{\tilde{e}_5^2}}^2}
\Big\{\ln{m_{_{\tilde{e}_5^1}}\over m_{_{\tilde{e}_5^2}}}+{A_{_{e_5}}(A_{_{e_5}}-\mu\cot\beta)
\over m_{_{\tilde{e}_5^1}}^2-m_{_{\tilde{e}_5^2}}^2}g(m_{_{\tilde{e}_5^1}},m_{_{\tilde{e}_5^2}})\Big\}
\;,\nonumber\\
%%%%%%%%%%%%%%%%%%%%%%%%%%%%%%%%%%%%%%%%%%%%%%%%%%%%%%%%%%
&&\Delta_{22}^{L}={G_{_F}m_{_{\nu_4}}^4\over\sqrt{2}\pi^2\sin^2\beta}\Big\{\ln{m_{_{\tilde{\nu}_4^1}}m_{_{\tilde{\nu}_4^2}}
\over m_{_{\nu_4}}^2}+{A_{_{\nu_4}}(A_{_{\nu_4}}-\mu\cot\beta)\over m_{_{\tilde{\nu}_4^1}}^2-m_{_{\tilde{\nu}_4^2}}^2}
\ln{m_{_{\tilde{\nu}_4^1}}^2\over m_{_{\tilde{\nu}_4^2}}^2}
\nonumber\\
&&\hspace{1.2cm}
+{A_{_{\nu_4}}^2(A_{_{\nu_4}}-\mu\cot\beta)^2\over(m_{_{\tilde{\nu}_4^1}}^2-m_{_{\tilde{\nu}_4^2}}^2)^2}
g(m_{_{\tilde{\nu}_4^1}}, m_{_{\tilde{\nu}_4^2}})\Big\}
\nonumber\\
&&\hspace{1.2cm}
+{G_{_F}m_{_{e_4}}^4\over\sqrt{2}\pi^2\cos^2\beta}\cdot
{\mu^2(A_{_{e_4}}-\mu\tan\beta)^2\over(m_{_{\tilde{e}_4^1}}^2-m_{_{\tilde{e}_4^2}}^2)^2}
g(m_{_{\tilde{e}_4^1}},m_{_{\tilde{e}_4^2}})
\nonumber\\
&&\hspace{1.2cm}
+{G_{_F}m_{_{\nu_5}}^4\over\sqrt{2}\pi^2\cos^2\beta}\cdot
{\mu^2(A_{_{\nu_5}}-\mu\tan\beta)^2\over(m_{_{\tilde{\nu}_5^1}}^2-m_{_{\tilde{\nu}_5^2}}^2)^2}
g(m_{_{\tilde{\nu}_5^1}},m_{_{\tilde{\nu}_5^2}})
\nonumber\\
&&\hspace{1.2cm}
+{G_{_F}m_{_{e_5}}^4\over\sqrt{2}\pi^2\sin^2\beta}\Big\{\ln{m_{_{\tilde{e}_5^1}}m_{_{\tilde{e}_5^2}}
\over m_{_{e_5}}^2}+{A_{_{e_5}}(A_{_{e_5}}-\mu\cot\beta)\over m_{_{\tilde{e}_5^1}}^2-m_{_{\tilde{e}_5^2}}^2}
\ln{m_{_{\tilde{e}_5^1}}^2\over m_{_{\tilde{e}_5^2}}^2}
\nonumber\\
&&\hspace{1.2cm}
+{A_{_{e_5}}^2(A_{_{e_5}}-\mu\cot\beta)^2\over(m_{_{\tilde{e}_5^1}}^2-m_{_{\tilde{e}_5^2}}^2)^2}
g(m_{_{\tilde{e}_5^1}}, m_{_{\tilde{e}_5^2}})\Big\}\;,
\label{app4-1}
\end{eqnarray}

\end{document}